\documentclass{new_tlp}
\usepackage{times,helvet,courier}
\usepackage{pdfsync} \synctex=1
\usepackage{todonotes} 
\usepackage{times}
\usepackage{verbatim} 
\usepackage{enumitem}
\usepackage{amsmath}
\usepackage{graphicx}
\usepackage{latexsym}
\usepackage{xspace}
\usepackage{url}
\newtheorem{lemma}{Lemma}

\newtheorem{theorem}{Theorem}

\newtheorem{definition}{Definition}

\newcommand{\classic}{\ensuremath{\mathit{cla}}}

\newcommand{\stable}{\ensuremath{\mathit{sta}}}
\newcommand{\atypeofmodel}{\ensuremath{w}}

\newcommand{\decisionsuperscript}{d}
\newcommand{\decision}[1]{#1^{\decisionsuperscript}}
\newcommand{\opposite}[1]{\overline{#1}}
\newcommand{\terminalstate}[1]{\mathit{Ok}(#1)}

\newcommand{\transitionarrow}{\Longrightarrow}

\newcommand{\aprogram}{\mathit{\Pi}}

\newcommand{\anatom}{a}
\newcommand{\anotheratom}{b}
\newcommand{\athirdatom}{c}
\newcommand{\afourthatom}{d}
\newcommand{\abody}{B}
\newcommand{\ahead}{A}
\newcommand{\aliteral}{l}

\newcommand{\astringofliterals}{L}

\newcommand{\asetofatoms}{X}

\newcommand{\asetofliterals}{M}
\newcommand{\amodel}{\asetofliterals}

\def\beq{\begin{equation}}
	\def\eeq#1{\label{#1}\end{equation}}
\def\ba{\begin{array}}
	\def\ea{\end{array}}

\newcommand{\logicaltrue}{\top}
\newcommand{\logicalfalse}{\bot}

\newcommand{\logicalnot}{\neg}

\newcommand{\setintersection}{\cap}
\newcommand{\setunion}{\cup}

\newcommand{\aspimplication}{\leftarrow}

\newcommand{\lexlower}{\leq_{\mathit{lex}}}

\newcommand{\atoms}[1]{{\mathit{atoms}(#1)}}

\newcommand{\reduct}[2]{#1^{#2}}

\newcommand{\depth}[1]{v(#1)}

\newcommand{\agraph}{G}

\newcommand{\overstablegraph}[1]{\mathit{OS}_{#1}}
\newcommand{\understablegraph}[1]{\mathit{US}_{#1}}
\newcommand{\mixedstablegraph}[1]{\mathit{MixS}_{#1}}
\newcommand{\chunkstablegraph}[1]{\mathit{CS}_{#1}}

\newcommand{\cmodels}{{\sc cmodels}\xspace}
\newcommand{\clasp}{{\sc clasp}\xspace}
\newcommand{\wasp}{{\sc wasp}\xspace}
\newcommand{\dlv}{{\sc dlv}\xspace}
\newcommand{\gnt}{{\sc gnt}\xspace}

\newcommand{\dpll}{{\sc dpll}\xspace}

\newcommand{\statesof}[1]{V_{#1}}

\newcommand{\literalsof}[1]{\mathit{lit}(#1)}
\newcommand{\overpropagators}{\mathit{ov}}
\newcommand{\underpropagators}{\mathit{un}}
\newcommand{\chunkpropagators}{\mathit{ch}}
\newcommand{\innerpropagators}{\mathit{in}}
\newcommand{\overapproxrule}{\mathit{OverApprox}}
\newcommand{\underapproxrule}{\mathit{UnderApprox}}
\newcommand{\chunkapproxrule}{\mathit{Chunk}}

\newcommand{\newsetrule}{\mathit{NewSet}}
\newcommand{\cautiousof}[1]{\mathit{cautious}(#1)}
\newcommand{\backboneof}[1]{\mathit{backbone}(#1)}

\newcommand{\anoverapproximation}{O}
\newcommand{\anunderapproximation}{U}
\newcommand{\foundornot}{f}

\newcommand{\afinalresult}{W}

\newcommand{\corestate}[5]{{#1}_{#2,#3,#4}}
\newcommand{\controlstate}[6]{{#1}({#3},{#4})}

\newcommand{\coreunderstate}[5]{{#1}_{#2,#3,#4}}
\newcommand{\controlunderstate}[6]{{#1}({#3},{#4})}

\newcommand{\continueinstruction}{\mathit{Cont}}

\newcommand{\prelfailinstruction}{\mathit{Pre}}

\newcommand{\succeedrule}{\mathit{Find}}

\newcommand{\failrule}{\mathit{Fail}}

\newcommand{\anaction}{A}
\newcommand{\overapproxaction}{\mathit{over}}
\newcommand{\underapproxaction}[1]{\mathit{under}_{#1}}
\newcommand{\chunkapproxaction}[1]{\mathit{chunk}_{#1}}
\newcommand{\initaction}{\mathit{chunk}}
\newcommand{\prelaction}[1]{\mathit{pre}_{#1}}
\newcommand{\achunk}{N}

\newcommand{\anaim}{T}

\usepackage{booktabs}
\usepackage{pgfplots}\usetikzlibrary{plotmarks}

\pgfplotsset{
	filter discard warning=false 
	, legend cell align=left
	, minor grid style={loosely dotted, lightgray}
	, major grid style={loosely dashed, lightgray}
}

\newtheorem{example}{Example}

\title{Abstract Solvers for Computing Cautious Consequences of ASP programs}

 \author[G. Amendola, C. Dodaro and M. Maratea]
 {GIOVANNI AMENDOLA, CARMINE DODARO\\
 University of Calabria, Italy\\
 \email{\{amendola,dodaro\}@mat.unical.it}
 \and
  MARCO MARATEA \\
 University of Genoa, Italy\\
 \email{marco@dibris.unige.it}
 }

 \jdate{March 2003}
\pubyear{2003}
\pagerange{\pageref{firstpage}--\pageref{lastpage}}
\doi{S1471068401001193}

\begin{document}

\label{firstpage}
\maketitle

\begin{abstract}
Abstract solvers are a method to formally analyze algorithms that have been profitably used for describing, comparing and composing solving techniques in various fields such as Propositional Satisfiability (SAT), Quantified SAT, Satisfiability Modulo Theories, Answer Set Programming (ASP), and Constraint ASP.

In this paper, we design, implement and test novel abstract solutions for cautious reasoning tasks in ASP. We show how to improve the current abstract solvers for cautious reasoning in ASP with new techniques borrowed from backbone computation in SAT, in order to design new solving algorithms. By doing so, we also formally show that the algorithms for solving cautious reasoning tasks in ASP are strongly related to those for computing backbones of Boolean formulas. We implement some of the new solutions in the ASP solver {\sc wasp} and show that their performance are comparable to state-of-the-art solutions on the benchmark problems from the past ASP Competitions.
Under consideration for acceptance in TPLP.
\end{abstract}

\begin{keywords}
 Answer Set Programming, Abstract solvers, Cautious reasoning
 \end{keywords}

\section{Introduction}

Abstract solvers are a  method to formally analyse solving
algorithms. In this methodology, the states of a computation are represented as
nodes of a graph, the solving techniques as edges between such nodes, the
solving process as a path in the graph,  and formal properties of the
algorithms are reduced to related graph properties. This framework enjoys some advantages w.r.t. traditional ways such as pseudo-code-based descriptions, e.g., being based on formal and well-known, yet simple, mathematical objects like graphs, which helps $(i)$ comparing solving algorithms by means of comparison of their related graphs, $(ii)$ mixing techniques in different algorithms in order to design novel (combination of) solving solutions, by means of mixing arcs in the related graphs, and $(iii)$ stating and proving formal properties of the solving algorithms, by means of reachability within the related graphs. Abstract solvers already proved to be a useful tool for formally describing, comparing and composing solving
techniques in various fields such as Propositional Satisfiability (SAT) and Satisfiability Modulo Theories (SMT)~\cite{nie06}, Quantified SAT~\cite{bro15}, Answer Set Programming~\cite{lier10,lier11,blm14}, and Constraint ASP~\cite{lie14}.
In ASP, such methodology led even to the development of a new ASP solver, {\sc sup}~\cite{lier10}; however, abstract solvers have been so far mainly applied to ASP solvers for brave reasoning tasks where, given an input query and a knowledge base expressed in ASP, answers are witnessed by ASP solutions, i.e., stable models~\cite{bar03,eite-etal-97f,gelf-lifs-88,gelf-lifs-91,mare-trus-98sm,nie99}.

However, in ASP, also cautious reasoning has been deeply studied in the literature: answers here must be witnessed by all stable models. This task has found a significant number of interesting applications as well, including consistent query answering~\cite{ArenasBC03,MannaRT13}, data integration~\cite{Eiter05}, multi-context systems~\cite{bre07}, and ontology-based reasoning~\cite{eit08}. Two well-known ASP solvers, i.e., {\dlv}~\cite{LeonePFEGPS06} and {\clasp}~\cite{GebserKS12}, have been extended for computing cautious consequences of ASP programs.
More recently, \citeN{AlvianoDR14} presented a unified, high-level view of such solving procedures, and designed
several algorithms for cautious reasoning in ASP, including those implemented in {\dlv} and {\clasp}, borrowed from the backbone computation of Boolean formulas~\cite{JanotaLM15}: all these techniques are implemented (and tested) on top of the ASP solver {\wasp}~\cite{alv15}.

In this paper we design, implement and test novel abstract solutions for cautious reasoning tasks in ASP. We show how to improve the current abstract solvers~\cite{bro15b} for cautious reasoning in ASP with further techniques borrowed from backbone computation in SAT, in order to design new solving algorithms. 
In particular, we import a technique called ``chunk'', which generalizes over- and under-approximation by testing a set soft atoms simultaneously for being added in the under-approximation, and core-based algorithms, which can be considered either a solution per se, or a way for pruning the set of atoms to be considered, given that they can not guarantee completeness.
By doing so, we also formally show, through a uniform treatment, that the algorithms for solving cautious reasoning tasks in ASP are strongly related to those for computing backbones of Boolean formulas. Finally, we implement some of the new solutions in the ASP solver \textsc{wasp}: results of a wide experimental analysis confirm that abstract solvers are a useful tool also for designing abstract solving procedures, given the performances of the related implementations are overall comparable to state-of-the-art solutions on the benchmark problems from the past ASP Competitions.

The paper is structured as follows. Section~\ref{sec:prel} introduces
needed preliminaries, including a review in  Section~\ref{sec:cm} of current 
algorithms for cautious reasoning trough abstract solving methodology.  Section~\ref{sec:backbone} shows
how the algorithms for computing backbones of Boolean formulas can be imported into ASP, to design new solving algorithms. It also contains a general theorem showing the relation between backbones computation in SAT and cautious reasoning in ASP.  Section~\ref{sec:exp} then presents the results of the new solutions on devoted ASP benchmarks. The paper ends by discussing
related work in Section~\ref{sec:related}, and by drawing conclusions in Section~\ref{sec:concl}.

\section{Preliminaries}
\label{sec:prel}

In this section, we first recall basics on (ground) non-disjunctive answer set programming (ASP) and Boolean logic formulas in Conjunctive Normal Form (CNF).
Then, we introduce the abstract solvers framework and its methodology.
Finally, we recall existing abstract solvers for computing cautious consequences of ASP programs.

\subsection{Boolean Formulas and Answer Set Programs}
We define (ground) non-disjunctive ASP programs and CNF formulas so as to underline similarities, in order to make it easier in later sections to compare algorithms working on CNF formulas with those working on ASP programs.

\paragraph{Syntax.}
Let $\Sigma$ be a propositional signature.
An element $a\in \Sigma$ is called \textit{atom} or \emph{positive literal}. The negation of an atom $a$, in symbols $\logicalnot\anatom$, is called \emph{negative literal}.
Given a literal $l$, we define $|l|=a$, if $l=a$ or $l=\neg a$, for some $a\in\Sigma$.
For a set of atoms $\asetofatoms\subseteq \Sigma$, a \textit{literal relative to $\asetofatoms$} is a
literal $l$ such that $|l|\in X$, and
$\literalsof{\asetofatoms}$ is the set of all literals relative to
$\asetofatoms$.
We set $\bar{l}=a$, if $l=\neg a$, and $\bar{l}=\neg a$, if $l=a$.
A \emph{clause} is a finite set of literals (seen as a disjunction).
A \emph{CNF formula} is a finite set of clauses (seen as a conjunction).
Given a set of literals $M$, we denote by $\asetofliterals^+$ the set of positive literals of
$\asetofliterals$, by $\asetofliterals^-$ the set of negative literals of $\asetofliterals$, and by $\overline{M}$ the set $\{\bar{l}\mid l\in M\}$.
We say that $M$ is \emph{consistent} if it does not contain
both a literal and its negation.
A (non-disjunctive) \textit{rule} is a pair $(\ahead,\abody)$, written
$\ahead\aspimplication\abody$, where $\abody$ is a finite set of literals and
$\ahead$ is an atom or the empty set.
We may write a rule as $\ahead\aspimplication\abody^+,\abody^-$,
as an abbreviation for $\ahead\aspimplication\abody^+\setunion\abody^-$, and
$\ahead\aspimplication\aliteral,\abody$ as an abbreviation for
$\ahead\aspimplication\{\aliteral\}\setunion\abody$.
A \textit{program} is a finite set of rules.
Given a set of literals $M$, a program $\Pi$, and a CNF formula $\Phi$, we denote by $\atoms{M}$, $\atoms{\aprogram}$, and $\atoms{\Phi}$ the set of
atoms occurring in $M$, $\aprogram$, and $\Phi$, respectively. 
It is important to emphasize here that the interpretation of negation is different in propositional formulas and in ASP programs. Indeed, in propositional formulas $\neg$ represents the classical negation, while in ASP programs it represents the \textit{negation by default}.

\paragraph{Semantics.}
An \emph{assignment} to a set $\asetofatoms$ of atoms is a total mapping
from $\asetofatoms$ to $\{\logicalfalse,\logicaltrue\}$.
We identify a consistent
set $\asetofliterals$ of literals with an assignment to
$\atoms{\asetofliterals}$ such that
$\anatom\in\asetofliterals$ iff $\anatom$ is mapped to $\logicaltrue$, and
$\logicalnot\anatom\in\asetofliterals$ iff $\anatom$ is mapped to
$\logicalfalse$.
A \emph{classical model} of a CNF formula $\Phi$ is an assignment $\asetofliterals$ to $\atoms{\Phi}$
such that for each clause $C \in \Phi$, $M \cap C \neq \emptyset$.
A \emph{classical model} of a program $\aprogram$
is an assignment $\asetofliterals$ to $\atoms{\aprogram}$ such that for
each rule $(A,B)$ $\in$ $\aprogram$, 
$A\cap M\neq \emptyset$ or $B\not\subseteq M$.
We denote  $M(\Phi)$ (resp. $M(\Pi)$) the set of all classical models of $\Phi$ (resp. $\Pi$).
The \emph{reduct} $\reduct{\aprogram}{\asetofatoms}$ of a program $\aprogram$
w.r.t. a set of atoms $\asetofatoms$ is obtained from $\aprogram$ by deleting
each rule $\ahead\aspimplication\abody^+,\abody^-$ such that
$\asetofatoms\setintersection\atoms{\abody^-}\neq\emptyset$
and replacing each remaining rule $\ahead\aspimplication\abody^+,\abody^-$ with
$\ahead\aspimplication\abody^+$.
An \emph{answer set} (or \textit{stable model}) of a program $\aprogram$ is an assignment
$\asetofliterals$ to $\atoms{\aprogram}$ such that
$\asetofliterals^+$ is minimal among the $\asetofliterals_0^+$ such that $\asetofliterals_0$
is a classical model of $\reduct{\aprogram}{\asetofliterals^+}$.
We denote by $AS(\Pi)$ the set of all answer sets of $\Pi$.
Given a formula $\Phi$ and a program $\Pi$, we define
$\backboneof{\Phi} = \bigcap_{M\in M(\Phi)} M^+$%
and
$\cautiousof{\aprogram} = \bigcap_{M\in AS(\Pi)} M^+$.

\begin{example}\label{ex:cautious}
	Consider the following program $\aprogram =\{
	\anatom  \aspimplication \logicalnot\anotheratom, \
	\anotheratom  \aspimplication \logicalnot\anatom, \
	\athirdatom  \aspimplication \anatom,\
	\athirdatom  \aspimplication \anotheratom\}$.
	$\Pi$ has two answer sets, namely $A_1=\{\neg a, b, c\}$ and $A_2=\{a,\neg b, c\}$.
		Hence, $A_1^+=\{b,c\}$ and $A_2^+=\{a,c\}$.
	Therefore, $\cautiousof{\aprogram} = \{b, c\} \cap \{a, c\} = \{c\}$.
	Now, consider the following CNF formula $\Phi = \{ a\vee b, \neg a\vee c, \neg b \vee c\}$. $\Phi$ has three classical models, namely $M_1=\{\neg a, b, c\}$, $M_2=\{a,\neg b, c\}$, and $M_3=\{a,b,c\}$.
		Hence, $M_1^+=\{b,c\}$, $M_2^+=\{a,c\}$, and $M_3^+=M_3$. 
		Therefore, $\backboneof{\Phi} = \{b, c\} \cap \{a, c\}\cap \{a,b,c\} = \{c\}$.
\end{example}

\subsection{Abstract Solvers for Solving CNF Formulas and ASP Programs}
\label{sec:as}
Now, we introduce the abstract solvers framework and its methodology employed later on in Section~\ref{sec:cm} and Section~\ref{sec:backbone} for computing cautious consequences of ASP programs. As we have mentioned in the introduction, abstract solvers are graphs that represent the status of the computation, and how it changes in response to an application of a technique in a search for a solution with certain properties, e.g., the satisfiability of a formula. Correspondingly, in the next paragraphs we first present the concept of a {\sl state}, i.e., all possible paths of the computation in terms of assignments, then the {\sl transition rules} are introduced, that showing how the state changes as a consequence of an application of a search technique if some conditions are met. The last paragraph of this subsection introduces {\sl abstract solver graphs}, where the states are the possible nodes of the graph, while transition rules define arcs among reachable nodes. \\

\paragraph{States.}
Given a set of atoms $\asetofatoms$, an \emph{action relative to $\asetofatoms$} is an element of the set $\mathcal{A}(X)=\{\overapproxaction,\mathit{under}_\emptyset\}\setunion \{\underapproxaction{\{a \}} \mid a\in\asetofatoms\}$.
For a set $\asetofatoms$ of atoms, a \emph{record} relative to~$\asetofatoms$ is a string $\astringofliterals$ from $\mathit{lit}(X)$ without repetitions.
A record $\astringofliterals$ is {\em consistent} if it does not contains both a literal and its negation.
We may view a record as the set containing all its elements stripped from their annotations.
For example, we may view $\logicalnot \anatom \anotheratom$ as $\{\logicalnot \anatom,\anotheratom\}$, and hence as the assignment that maps $\anatom$ to $\logicalfalse$ and $\anotheratom$ to $\logicaltrue$.
Given a set $X$ of atoms, the set of \emph{states relative to $\asetofatoms$}, written
$\statesof{\asetofatoms}$, is the union of:
\begin{itemize}
	\item[$(i)$] the set of \emph{core states relative to $\asetofatoms$}, that are all $\corestate{\astringofliterals} {\anoverapproximation}{\anunderapproximation}{\anaction}{\foundornot}$ such that $\astringofliterals$ is a record relative to $\asetofatoms$; 
	$\anoverapproximation$, $\anunderapproximation$ $\in\asetofatoms$; and
	$\anaction \in \mathcal{A}(\asetofatoms)$;
	\item[$(ii)$] the set of \emph{control states relative to $\asetofatoms$}, that are all the $\controlstate{\continueinstruction}{\astringofliterals} {\anoverapproximation}{\anunderapproximation}{\anaction}{\foundornot}$ where $\anoverapproximation$, $\anunderapproximation$ $\in \asetofatoms$;
	and \item[$(iii)$] the set of \emph{terminal states relative to $\asetofatoms$}, that are all $\terminalstate{\afinalresult}$, where $\afinalresult \in \asetofatoms$. 
\end{itemize}

Intuitively, these states represent computation steps of the algorithms that search for assignments with certain properties, in our case being backbone or cautious consequence.
The computation starts from a specific core state, called \textit{initial state}, depending on the specific algorithm (concrete examples are given later when presenting the techniques).
Other core states $\corestate{\astringofliterals} {\anoverapproximation}{\anunderapproximation}{\anaction}{\foundornot}$
and the control states $\controlstate{\continueinstruction}{\astringofliterals} {\anoverapproximation}{\anunderapproximation}{\anaction}{\foundornot}$ represent all the intermediate steps of the computation, where $\astringofliterals$ is the current state of the computation of a model;
$\anoverapproximation$ is the current over-approximation of the solution; $\anunderapproximation$ is the current under-approximation of the solution; and $\anaction$ is the action currently carried out: 
$\overapproxaction$ (resp. $\mathit{under}_\emptyset$ or $\underapproxaction{\{a\}}$) if over-approximation (resp. under-approximation) is being applied.
Intuitively, a core state represents the computation within a call to an ASP oracle, i.e., an ASP solver, while a control state controls the computation between different calls to ASP oracles, depending on over-approximation and under-approximation. The terminal states represent the end of the computation, i.e., the termination of the algorithm.

For instance, consider the following set of atoms $X=\{a,b,c\}$. Hence, $lit(X)=\{a,b,c,\neg a,$ $\neg b, \neg c\}$. 
Therefore, $\neg a b_{\{a,b\},\emptyset,\mathit{over}}$ is an example of core state relative to $X$ where $a$ is assigned to false and $b$ to true, the over-approximation is the set $\{a,b\}$ while the under-approximation is empty, and the action executed is over. Other examples of core states are $\emptyset_{\{a\},\{b\},\mathit{under}_\emptyset}$ and
$\neg a\neg b\neg c_{\emptyset,\emptyset,\mathit{under}_{\{ a\}}}$.
Instead, $\mathit{Cont}(\{a,b\},\{a\})$, $\mathit{Cont}(\{a,b,c\},\emptyset)$, $\mathit{Cont}(\emptyset,\emptyset)$ are examples of control states relative to $X$, where e.g., in the first example the over-approximation is the set $\{a,b\}$ and the under-approximation is $\{a\}$. 
$\mathit{Ok}(\{a,b,c\})$ and $\mathit{Ok}(\emptyset)$ are examples of terminal states relative to $X$, where set $\{a,b,c\}$ and $\emptyset$ are solutions.

\paragraph{Transition Rules.}
\textit{Transition rules} are represented with the following structure:
$$
\begin{array}{llll}
ruleName & S & \transitionarrow \ S' & \textrm{if}\left\{\ conditions\right.
\end{array}
$$
where, 
$(i)$ $ruleName$ is the name of the rule; 
$(ii)$ $S \transitionarrow S'$ represents a transition from the starting state $S$ to the arriving state $S'$ (if the rule is applied); 
and $(iii)$ $conditions$ is a set of conditions for the rule to be applicable.

We also consider a special transition rule, called $\mathit{Oracle}$, which starts from a state $L_{O,U,A}$ and arrives to a  state $L'_{O,U,A}$, if $L=\emptyset$. In symbols:
$$
\begin{array}{llll}
\mathit{Oracle}
& \corestate {\astringofliterals} {\anoverapproximation}{\anunderapproximation} {\anaction}{\foundornot}
& \transitionarrow \ \corestate {\astringofliterals'} {\anoverapproximation}{\anunderapproximation} {\anaction}{\foundornot}
& \textrm{if}\left\{\ \astringofliterals = \emptyset \right.
\end{array}
$$
Intuitively, the $\mathit{Oracle}$ rule represents an oracle call to an ASP [resp., SAT] solver by providing as result a set of literals $L'$ corresponding to the output of an ASP [resp., SAT] solver, i.e., $L'$ will correspond to an answer set of a logic program [resp., a classical model of a Boolean formula], if such an answer set [resp. classical model] exists, and to an inconsistent set of literals, otherwise. Transition rules in our paper are organized into $\mathit{Return}$ and ${Control}$ rules. Return rules deal with the outcome of an oracle call, or the application of a given technique, depending on the status of the set of literals $L$ returned, while Control rules start from a control state an direct the computation depending on the content of the over- and under-approximation.

\paragraph{Abstract Solver Graphs.}
Given a set of atoms $X$ and a set of transition rules $T$, we define an \textit{abstract solver graph} $G_{X,T}=\langle V_X, E_{T}\rangle$, where $(S,S')\in E_{T}$ if, and only if, a transition rule of the form  $S \transitionarrow S'$ can be applied.
We also denote the set of edges $E_{T}$ by the set of transition rules $T$. 
We say that a state $S\in V_X$ is \textit{reachable from} a state $S'\in V_X$, if there is a path from $S'$ to $S$.
Every state reachable from the initial state is called \textit{reachable state}, and represents a possible state of a computation. 
Each path starting from the initial state represents the description of possible search for a certain model.
We say that \emph{no cycle is reachable} if there is no
reachable state which is reachable from itself.
Finally, note that transition rules $T$ and the set $X$ will depend from the specific input program $\Pi$, thus instead of writing $G_{X,T}$, we will write just $G_\Pi$. 

\subsection{Naive Abstract Solvers for Computing Cautious Consequences}
\label{sec:cm}
In this section, we recall the abstract \textit{over-approximation}, \textit{under-approximation} and \textit{mixed} strategies
for computing cautious consequences of ASP programs. 

\begin{definition}
	Given a program $\Pi$ [resp., a CNF formula $\Phi$], we say that an abstract solver graph $\agraph_\Pi$ [resp., $\agraph_\Phi$] \textit{solves cautious reasoning} [resp., backbone computation],  if
	$(i)$ $\agraph_\Pi$  [resp., $\agraph_\Phi$] is finite and no cycle is reachable; and
	$(ii)$ the unique terminal reachable state in $\agraph_\Pi$  [resp., $\agraph_\Phi$] is 
	$Ok(cautious(\Pi))$ [resp., $Ok(backbone(\Pi))$].	
\end{definition}
In the following, without loss of generality, we only focus on the computation of cautious consequences for an ASP program $\Pi$.

\paragraph{General Structure.}
Given a program $\Pi$, over-approximation is set to all atoms in the program, i.e., $O=\atoms{\Pi}$, while the under-approximation is empty, i.e., $U=\emptyset$.  Note that $U\subseteq \mathit{cautious}(\Pi)\subseteq O$.
Iteratively either under-approximation or over-approximation are applied.
When they coincide, i.e., $U=O$, 
the set of cautious consequences, i.e., $\anoverapproximation$, has been found and the computation terminates. 
It means that the state $\terminalstate{\anoverapproximation}$ is a reachable state.
Hence, the full extent of states relative to $\asetofatoms$ becomes useful.
The unique terminal state is $\terminalstate{\afinalresult}$, where $\afinalresult$ is the set of all cautious consequences of $\Pi$.

\paragraph{Over-approximation.}
\begin{figure}[t] \footnotesize{
		$$
		\arraycolsep=3pt
		\begin{array}{llll}
		\multicolumn{4}{l}{\textrm{Return rules}}\\
		
		\failrule_{\overapproxaction}
		& \corestate
		{\astringofliterals}
		{\anoverapproximation}{\anunderapproximation}
		{\overapproxaction}{\logicaltrue}
		& \transitionarrow
		\controlstate
		{\continueinstruction}
		{\astringofliterals}
		{\anoverapproximation}{\anoverapproximation}
		{\anaction}{\logicaltrue}
		& \textrm{if}\left\{
		\begin{array}{l}
			\astringofliterals\textrm{ is inconsistent}
		\end{array}
		\right.\\
		
		\succeedrule
		& \corestate
		{\astringofliterals}
		{\anoverapproximation}{\anunderapproximation}
		{\anaction}{\foundornot}
		& \transitionarrow
		\controlstate
		{\continueinstruction}
		{\astringofliterals}
		{\anoverapproximation\setintersection\astringofliterals}{\anunderapproximation}
		{\anaction}{\logicaltrue}
		& \textrm{if}\left\{
		\begin{array}{l}
			L\textrm{ is consistent and } L\neq\emptyset
		\end{array}
		\right.\\
		\\
		
		\multicolumn{4}{l}{\textrm{Control rules}}\\
		\mathit{Terminal}
		& \controlstate
		{\continueinstruction}
		{\astringofliterals}
		{\anoverapproximation}{U}
		{\anaction}{\foundornot}
		& \transitionarrow
		\terminalstate{\anoverapproximation}&
		\textrm{if}\left\{
		\begin{array}{l}
			\anoverapproximation=\anunderapproximation
		\end{array}
		\right.\\
		
		\overapproxrule
		& \controlstate
		{\continueinstruction}
		{\astringofliterals}
		{\anoverapproximation}{\anunderapproximation}
		{\anaction}{\foundornot}
		& \transitionarrow
		\corestate
		{\emptyset}
		{\anoverapproximation}{\anunderapproximation}
		{\overapproxaction}{\logicaltrue}
		& \textrm{if}\left\{
		\begin{array}{l}
			\anoverapproximation\neq\anunderapproximation
		\end{array}
		\right.\\
	\end{array}
	$$
}
\normalsize
\caption{The transition rules of $\overpropagators$.}
\label{fig:trover}
\end{figure}

Let
$\aprogram_{\anoverapproximation,\anunderapproximation,\overapproxaction}$
$=$
$\aprogram
\setunion\{\aspimplication\anoverapproximation\}$.
The initial state is
$\corestate{\emptyset}
{\atoms{\aprogram}}{\emptyset}
{\mathit{over}}{\logicalfalse}$. 
We call $\overpropagators$ the set of all the rules reported in
Figure~\ref{fig:trover}, that is
$\overpropagators=\{\failrule_{\overapproxaction},\linebreak[1]\succeedrule,\linebreak[1]\mathit{Terminal},\linebreak[1]\overapproxrule\}$.
Intuitively,
$\failrule_{\overapproxaction}$ means that a call to an oracle did not find an answer set, so $\anoverapproximation$ is the solution. If $\succeedrule$ is
triggered, instead, we go to a control state where $\anoverapproximation$ is updated according to the answer set found: then, if
$\anoverapproximation=\anunderapproximation$ a solution is found through
$\mathit{Terminal}$, otherwise the search is restarted
($\astringofliterals=\emptyset$) in an oracle state with $\overapproxrule$. For
any $\aprogram$, the graph $\overstablegraph{\aprogram}$ is
$(\statesof{\atoms{\aprogram}},\{\mathit{Oracle}\}\setunion\overpropagators)$.
Thus, in $\overstablegraph{\aprogram}$, the oracle is called to
find answer sets that reduce the over-approximation
$\anoverapproximation$ in the $\overapproxaction$ action, unless no answer set exists. If an answer set $\asetofliterals$ is
found, then $\asetofliterals\setintersection\overline{\anoverapproximation}\neq\emptyset$, as $\aprogram_{\anoverapproximation,\anunderapproximation,\overapproxaction}$
$=$ $\aprogram
\setunion\{\aspimplication\anoverapproximation\}$. 

Indeed, assume by contradiction that $M\cap \overline{\anoverapproximation}=\emptyset$, then $O\subseteq M$.
Hence, $M$ is not a model of the rule $(\emptyset, O)$, as $M\cap \emptyset =\emptyset$ and $O\subseteq M$.
Therefore, $M$ should not be a model of $\aprogram_{\anoverapproximation,\anunderapproximation,\overapproxaction}$, against the assumption that $M$ is an answer set of $\aprogram_{\anoverapproximation,\anunderapproximation,\overapproxaction}$.

\paragraph{Under-approximation.}
Let
$\aprogram_{\anoverapproximation,\anunderapproximation,\underapproxaction{\{a\}}}$
$=$ $\aprogram\setunion\{\aspimplication a\}$ and $\aprogram_{\anoverapproximation,\anunderapproximation,\underapproxaction{\emptyset}}$ $=$ $\aprogram$.
The initial state is
$\corestate{\emptyset}
{\atoms{\aprogram}}{\emptyset}
{\mathit{under}_\emptyset}{\logicalfalse}$.
We call $\underpropagators$ the set
$\{\failrule_{\underapproxaction{}},\linebreak[1]\succeedrule,\linebreak[1]\mathit{Terminal},\linebreak[1]\underapproxrule\}$
containing the rules presented in Figure~\ref{fig:trunder} plus
$\succeedrule$ and $\mathit{Terminal}$ from Figure~\ref{fig:trover}. Intuitively, 
$\failrule_{\underapproxaction{}}$ updates over- and under-approximations in case a test on the atom $a$ failed, and leads to a control state, while $\underapproxrule$ restarts a new test if $\succeedrule$ is not applicable.
For any $\aprogram$, 
the graph $\understablegraph{\aprogram}$ is
$(\statesof{\atoms{\aprogram}},\{\mathit{Oracle}\}\cup\underpropagators)$.
In $\understablegraph{\aprogram}$, again, a first oracle call takes place with
the action $\mathit{under}_\emptyset$, which provides first over-approximation, then calls with actions $\underapproxaction{\{a\}}$, where the element $a$ is the tested atom.
Figure~\ref{fig:example2-under} shows a possible path in $\understablegraph{\aprogram}$ for the program $\Pi$ of Example~\ref{ex:cautious}.
For compactness, the syntax in which the path is presented is slighly different, with ``$\transitionarrow$'' replaced by ``:'', and with the initial state not explicitly tagged. 

\begin{figure}[t] \footnotesize{
		$$
		\arraycolsep=3pt
		\begin{array}{llll}
		\multicolumn{4}{l}{\textrm{Return rule}}\\
		\failrule_{\underapproxaction{}}
		& \corestate
		{\astringofliterals}
		{\anoverapproximation}{\anunderapproximation}
		{\underapproxaction{S}}{\foundornot}
		& \transitionarrow
		\controlstate
		{\continueinstruction}
		{\astringofliterals}
		{\anoverapproximation}{\anunderapproximation\setunion S}
		{\underapproxaction{S}}{\foundornot}
		& \textrm{if}\left\{
		\begin{array}{l}
		\astringofliterals\textrm{ is inconsistent, and } S=\emptyset \mbox{ or } S=\{a\}
		\end{array}
		\right.\\
		\\
		
		\multicolumn{4}{l}{\textrm{Control rule}}\\
		\underapproxrule
		& \controlstate
		{\continueinstruction}
		{\astringofliterals}
		{\anoverapproximation}{\anunderapproximation}
		{\anaction}{\foundornot}
		& \transitionarrow
		\corestate
		{\emptyset}
		{\anoverapproximation}{\anunderapproximation}
		{\underapproxaction{\{a \}}}{\logicaltrue}
		& \textrm{if}\left\{
		\begin{array}{l}
			a\in\anoverapproximation\setminus\anunderapproximation
		\end{array}
		\right.\\
	\end{array}
	$$
}
\normalsize
\caption{The transition rules of $\underpropagators$ that are not in
	$\overpropagators$.}
\label{fig:trunder}
\end{figure}

\begin{figure}[t]
	\footnotesize{
		$$
		\arraycolsep=3pt
		\begin{array}{l|l}
		\begin{array}{ll}
		\multicolumn{2}{l}{  \aprogram 
			= \aprogram_{
				\{\anatom,\anotheratom,\athirdatom\},\emptyset,
				\mathit{under}_\emptyset
			}
			= \left\{
			\begin{array}{l}
			\anatom  \aspimplication \logicalnot\anotheratom\\
			\anotheratom  \aspimplication \logicalnot\anatom\\
			\athirdatom  \aspimplication \anatom\\
			\athirdatom  \aspimplication \anotheratom\ 
			\end{array}\right\} }\\
		\\
		\\
		\\
		\multicolumn{2}{l}{
			\aprogram_{
				\{a,\athirdatom\},\emptyset,
				\underapproxaction{\{\athirdatom\}}
			}
			= \aprogram\setunion\{
			\aspimplication \athirdatom
			\} }\\
		\\
		\vspace{4pt}
		\\
		\multicolumn{2}{l}{
			\aprogram_{
				\{a,\athirdatom\},\{\athirdatom\},
				\underapproxaction{\{\anatom\}}
			}
			= \aprogram\setunion\{
			\aspimplication \anatom
			\} }\\
		\\
		\\
		\end{array}
		
		&
		
		\begin{array}{ll}
			&
			\corestate
			{\emptyset}
			{\{\anatom,\anotheratom,\athirdatom\}}{\emptyset}
			{\mathit{under}_\emptyset}{\logicaltrue}
			\\
			\mathit{Oracle}\mbox{ :}
			&
			\corestate
			{\anatom\athirdatom{\logicalnot\anotheratom}}
			{\{\anatom,\anotheratom,\athirdatom\}}{\emptyset}
			{\mathit{under}_\emptyset}{\logicaltrue}
			\\
			\succeedrule:
			&
			\controlstate
			{\continueinstruction}
			{\decision{\anatom}\athirdatom{\logicalnot\anotheratom}}
			{\{\anatom,\athirdatom\}}{\emptyset}
			{\initaction}{\logicaltrue}
			\\
			\\

			\underapproxrule:
			&
			\corestate
			{\emptyset}
			{\{\anatom,\athirdatom\}}{\emptyset}
			{\underapproxaction{\{\athirdatom\}}}{\logicaltrue}
			\\
			
			\mathit{Oracle}\mbox{ :}
			&
			\corestate
			{\logicalnot{\athirdatom}\logicalnot{\anatom}\anotheratom\athirdatom}
			{\{\anatom,\athirdatom\}}{\emptyset}
			{\underapproxaction{\{\athirdatom\}}}{\logicaltrue}
			\\

			\failrule_{\underapproxaction{}}:
			&
			\controlstate
			{\continueinstruction}
			{\logicalnot{\athirdatom}\decision{\anatom}\athirdatom}
			{\{\anatom,\athirdatom\}}{\{\athirdatom\}}
			{\underapproxaction{\{\athirdatom\}}}{\logicaltrue}
			\\
			\\
			
			\underapproxrule:
			&
			\corestate
			{\emptyset}
			{\{\anatom,\athirdatom\}}{\{\athirdatom\}}
			{\underapproxaction{\{\anatom\}}}{\logicaltrue}
			\\
			\mathit{Oracle}\mbox{ :}
			&
			\corestate
			{\logicalnot{\anatom}\anotheratom\athirdatom}
			{\{\anatom,\athirdatom\}}{\{\athirdatom\}}
			{\underapproxaction{\{\anatom\}}}{\logicaltrue}
			\\
			\succeedrule:
			&
			\controlstate
			{\continueinstruction}
			{\decision{\anatom}\athirdatom{\logicalnot\anotheratom}}
			{\{\athirdatom\}}{\{\athirdatom\}}
			{\underapproxaction{\{\anatom\}}}{\logicaltrue}
			\\
			\mathit{Terminal}:
			&
			\terminalstate{\{\athirdatom\}}
		\end{array}
	\end{array}
	$$
}
\caption{A path in $\understablegraph{\aprogram}$.}
\label{fig:example2-under}
\end{figure}

\paragraph{Mixed strategy.}
An abstract mixed strategy can be obtained by defining $\mixedstablegraph{\aprogram}$ as
$(\statesof{\atoms{\aprogram}},$ $\{\mathit{Oracle}\}\setunion\underpropagators\setunion\overpropagators)$.
Therefore, it is possible to combine techniques described by the graph for over-approximation and those in the graph for under-approximation, by envisaging the design of new additional algorithms.
Here, we have two potential initial states, i.e., $\emptyset_{\mathit{atoms}(\Pi),\emptyset,A}$, where $A\in\{\mathit{over},\mathit{under}_\emptyset\}$, i.e., depending whether over-appoximation or under-approximation is first applied. 

\section{Advanced Abstract Solvers for Computing Cautious Consequences}
\label{sec:backbone}
In this section we import in ASP further algorithms from \cite{JanotaLM15} through abstract solvers. 
First, we generalize the concepts of under- and over-approximation via chunks, which consider a set of atoms simultaneously. 
Then, we model core-based algorithms.
Finally, we state a general theorem, which includes all previous results, that shows how the techniques presented can be combined to design new solving methods for finding cautious consequences of ASP programs, and states a strong analogy between algorithms for computing cautious consequences of ASP programs and those for backbones of CNF formulas.

The sets of states now include also the following:
$\{\chunkapproxaction{\achunk} | \achunk\subseteq\atoms{\aprogram}\}$, ${\initaction}$ and 
$\{\mathit{core}_N | N\subseteq lit(\atoms{\aprogram})\}$.

\begin{figure}[t] \footnotesize{
		$$
		\arraycolsep=3pt
		\begin{array}{llll}
		\multicolumn{4}{l}{\textrm{Return rules}}\\
		\failrule_{\chunkapproxaction{}}
		& \coreunderstate
		{\astringofliterals}
		{\anoverapproximation}{\anunderapproximation}
		{\chunkapproxaction{\achunk}}{\foundornot}
		& \transitionarrow
		\controlunderstate
		{\continueinstruction}
		{\astringofliterals}
		{\anoverapproximation\setminus \achunk}{\anunderapproximation\setunion\achunk}
		{\anaction}{\foundornot}
		& \textrm{if}\left\{
		\begin{array}{l}
		\astringofliterals\textrm{ is inconsistent}
		\end{array}
		\right.\\
		\\
		
		\multicolumn{4}{l}{\textrm{Control rules}}\\
		\chunkapproxrule
		& \controlunderstate
		{\continueinstruction}
		{\astringofliterals}
		{\anoverapproximation}{\anunderapproximation}
		{\anaction}{\foundornot}
		& \transitionarrow
		\coreunderstate
		{\emptyset}
		{\anoverapproximation}{\anunderapproximation}
		{\chunkapproxaction{\achunk}}{\foundornot}
		& \textrm{if}\left\{
		\begin{array}{l}
			\achunk\subseteq\anoverapproximation\setminus\anunderapproximation
			\textrm{ and }
			\achunk\neq\emptyset
		\end{array}
		\right.\\
	\end{array}
	$$
}
\caption{The transition rules of $\chunkpropagators$ that are not in $\mathit{ov}$.}
\label{fig:classchunk}
\end{figure}
\normalsize

\subsection{Chunking}
\label{sec:ch}
In~\cite{JanotaLM15} a more general technique for
under-approximation that allows to test multiple literals at once is presented (see, Algorithm 5 in~\cite{JanotaLM15}).
We define $\chunkpropagators$ as the set
$\{\failrule_{\chunkapproxaction{}},\linebreak[1]\succeedrule,\linebreak[1]\mathit{Terminal},\linebreak[1]\chunkapproxrule\}$
containing the rules presented in Figure~\ref{fig:classchunk} plus
$\succeedrule$ and $\mathit{Terminal}$ from Figure~\ref{fig:trover}. The newly
introduced rules in Figure~\ref{fig:classchunk} model the new technique. In
particular, $\failrule_{\chunkapproxaction{}}$ updates the over- and
under-approximations accordingly in case the test
on the set $\achunk$ fails (the ASP oracle call failed, thus all literals in $N$ must be cautious consequences), and goes to a control state. Meanwhile,
$\chunkapproxrule$ restarts a new ASP oracle call with a new (nonempty) set $N$
such that $\achunk\subseteq\anoverapproximation\setminus\anunderapproximation$ in case the computation must continue (cf. condition of this transition rule). 
For any $\aprogram$, the graph $\mathit{CS}_\aprogram$ is
$(\statesof{\atoms{\aprogram}},\{\mathit{Oracle}\}\setunion\chunkpropagators)$.  The initial state is
$\corestate{\emptyset}
{\atoms{\aprogram}}{\emptyset}
{\initaction}{\logicaltrue}$.
We define
$\aprogram_{\anoverapproximation,\anunderapproximation,\chunkapproxaction{\achunk}}$
as $\aprogram\setunion\{\aspimplication\achunk\}$.

\begin{theorem}~\label{prop:chunkclass}
	Let $\Pi$ be a program. Then, the graph $\mathit{CS}_\aprogram$ solves cautious reasoning.
\end{theorem}

\begin{figure}[t]
	\footnotesize{
		$$
		\arraycolsep=3pt
		\begin{array}{l|l}
		\begin{array}{ll}
		\multicolumn{2}{l}{  \aprogram 
			= \aprogram_{
				\{\anatom,\anotheratom,\athirdatom,\afourthatom\},\emptyset,
				\initaction
			}
			= \left\{
			\begin{array}{l}
			\anatom  \aspimplication \logicalnot\anotheratom\\
			\anotheratom  \aspimplication \logicalnot\anatom\\
			\athirdatom  \aspimplication \anatom\\
			\athirdatom  \aspimplication \anotheratom\\ 
			\afourthatom  \aspimplication \athirdatom
			\end{array}\right\} }\\
		\\
		\\
		
		\multicolumn{2}{l}{
			\aprogram_{
				\{\anatom,\athirdatom,\afourthatom\},\emptyset,
				\chunkapproxaction{\{\athirdatom,\afourthatom\}}
			}
			= \aprogram\setunion\{
			\aspimplication \athirdatom, \afourthatom\ \}}
		\vspace{1cm}
		\\
		\\
		
		\multicolumn{2}{l}{
			\aprogram_{
				\{\anatom,\athirdatom,\afourthatom\},\{\athirdatom,\afourthatom\},
				\chunkapproxaction{\{\anatom\}}
			}
			= \aprogram\setunion\{
			\aspimplication \anatom
			\}}
		\\
		\\
		
		\\
	\end{array}
	
	&
	
	\begin{array}{ll}
		& \corestate
		{\emptyset}
		{\{\anatom,\anotheratom,\athirdatom,\afourthatom\}}{\emptyset}
		{\initaction}{\logicaltrue}
		
		\\
		\mathit{Oracle}\mbox{ :}
		&
		\corestate
		{\anatom\athirdatom\logicalnot\anotheratom\afourthatom}
		{\{\anatom,\anotheratom,\athirdatom,\afourthatom\}}{\emptyset}
		{\initaction}{\logicaltrue}
		\\
		\succeedrule:
		&
		\controlstate
		{\continueinstruction}
		{\athirdatom\decision{\anatom}\logicalnot\anotheratom}
		{\{\anatom,\athirdatom,\afourthatom\}}{\emptyset}
		{\initaction}{\logicaltrue}
		\\
		\\
		
		\chunkapproxrule:
		&
		\corestate
		{\emptyset}
		{\{\anatom,\athirdatom,\afourthatom\}}{\emptyset}
		{\chunkapproxaction{\{\athirdatom,\afourthatom\}}}{\logicaltrue}
		\\
		\mathit{Oracle}\mbox{ :}
		&
		\corestate
		{\logicalnot{\anatom}\anotheratom\athirdatom\afourthatom\logicalnot{\afourthatom}}
		{\{\anatom,\athirdatom,\afourthatom\}}{\emptyset}
		{\chunkapproxaction{\{\athirdatom,\afourthatom\}}}{\logicaltrue}
		\\
		\failrule_{\chunkapproxaction{}}: 
		&
		\controlstate
		{\continueinstruction}
		{\athirdatom\logicalnot\anatom\anotheratom}
		{\{\anatom,\athirdatom,\afourthatom\}}{\{\athirdatom,\afourthatom\}}
		{\overapproxaction}{\logicaltrue}
		\\
		\\
		
		\chunkapproxrule:
		&
		\corestate
		{\emptyset}
		{\{\anatom,\athirdatom,\afourthatom\}}{\{\athirdatom,\afourthatom\}}
		{\chunkapproxaction{\{\anatom\}}}{\logicaltrue}
		\\
		\mathit{Oracle}\mbox{ :}
		&
		\corestate
		{\logicalnot{\anatom}\anotheratom\athirdatom\afourthatom}
		{\{\anatom,\athirdatom,\afourthatom\}}{\{\athirdatom,\afourthatom\}}
		{\chunkapproxaction{\{\anatom\}}}{\logicaltrue}
		\\
		
		\succeedrule:
		&
		\controlstate
		{\continueinstruction}
		{\logicalnot{\athirdatom}\decision{\anatom}\athirdatom}
		{\{\athirdatom,\afourthatom\}}{\{\athirdatom,\afourthatom\}}
		{\chunkapproxaction{\{\anatom\}}}{\logicaltrue}
		\\
		\mathit{Terminal}:
		&
		\terminalstate{\{\athirdatom,\afourthatom\}}
		\\
		\\
		
	\end{array}
\end{array}
$$
}
\caption{A path in $\chunkstablegraph{\aprogram}$.}
\label{fig:example-chunk}
\end{figure}

In order to design a meaningful example of Chunk, we slightly modify our running example adding the rule $d \leftarrow c$. Figure~\ref{fig:example-chunk} shows a possible path in $\mathit{CS}_\aprogram$ for the new defined program.

\subsection{Designing New Abstract Solvers}

The composition of techniques described in Section \ref{sec:cm} and \ref{sec:ch} can be
readily applied to computing cautious consequences of a program, but
actually is not included in any solver. This outlines another important feature
of the abstract solvers methodology, i.e., its capability to design new solutions by means of combination of techniques implemented in different solvers.

More generally, it is possible to mix under-approximation, over-approximation, and chunking
technique, and apply them for computing
either cautious consequences or backbones. We next state a general
theorem that subsumes all the techniques previously described, showing a strong analogy among the algorithms for computing cautious consequences and those for backbones.

\begin{theorem}~\label{prop:general}
	For any program $\aprogram$, and for any set
	$S\subseteq\{\underpropagators,\overpropagators,\chunkpropagators\}$ such that
	$S\neq\emptyset$,
	the graph
	$(\statesof{\atoms{\aprogram}},$
	$\{\mathit{Oracle_{ASP}}\}
	\setunion\bigcup_{x\in S}x
	)$
	solves cautious reasoning, and
	the graph
	$(\statesof{\atoms{\aprogram}},$
	$\{\mathit{Oracle_{SAT}}\}
	\setunion\bigcup_{x\in S}x
	)$
	solves backbone computation, where $\mathit{Oracle_{ASP}}$ and $\mathit{Oracle_{SAT}}$  represent an oracle call to an ASP solver and to a SAT solver, respectively. 	
\end{theorem}

\subsection{Core-based Methods}

We now model core-based algorithms from~\cite{JanotaLM15} in terms of abstract solvers, in particular Algorithm 6, and apply it to the computation of cautious consequences of ASP programs. 
First, note that
$\Pi_{\anoverapproximation,\anunderapproximation,\mathit{core}_{\achunk}}$
is
$\Pi\setunion\{\aspimplication\overline{\aliteral}|\aliteral\in\achunk\}$,
and $\emptyset_{atoms(\Pi),\emptyset,\mathit{core}_{\overline{atoms(\Pi)}}}$ is the initial state.
Moreover, given a logic program $\Pi$, we say that a set $C\subseteq lit(atoms(\Pi))$ is \textit{a core} of $\Pi$, if $\Pi \cup \{\leftarrow \overline{l} | l\in C\}$ is incoherent.
It is important to emphasize here that this definition is in line with the one proposed by \citeN{AlvianoDJMP18}. In particular, unsatisfiable cores have two important properties:
	\begin{itemize}
		\item if $C$ is an unsatisfiable core of $\Pi$ then all of its supersets are also unsatisfiable cores of $\Pi$;
		\item an atom $p \in atoms(\Pi)$ is a cautious consequence of $\Pi$ if and only if $\{\neg p\}$ is an unsatisfiable core (Proposition 4.1 of \cite{AlvianoDJMP18}).
	\end{itemize}
	Moreover, in general unsatisfiable cores are not guaranteed to be minimal, albeit several strategies can be used to obtain a minimal unsatisfiable core~\cite{LynceM04,DBLP:journals/tplp/AlvianoD16,AlvianoDJMP18}. 
\begin{example}
	Consider the program $\Pi$ of the Example~\ref{ex:cautious} and let $N=\{\lnot a,\lnot b,\lnot c\}$. Hence, $\{\lnot c\}$, $\{\lnot a,\lnot c\}$, $\{\lnot b,\lnot c\}$, $\{\lnot a,\lnot b\}$, and $\{\lnot a,\lnot b,\lnot c\}$ are all cores of $\Pi_{O,U,\mathit{core}_N}$.
\end{example}

First, we consider a transition rule, called $\mathit{CoreOracle}$, which starts from a state $\emptyset_{O,U,\mathit{core}_N}$ and arrives to a  state $L'_{O,U,\mathit{core}_N}$. In symbols:
$$
\begin{array}{llll}
\mathit{CoreOracle}
& L_{O,U,\mathit{core}_N}
& \transitionarrow \
L'_{O,U,\mathit{core}_N}
& \textrm{if}\left\{\ L =\emptyset \right.
\end{array}
$$
The $\mathit{CoreOracle}$ rule represents an oracle call to compute a set of literals $L'$, which is 
an inconsistent set of literals such that the set $\widehat{L'}=\{\neg a\mid \{a,\neg a\}\subseteq L'\} $ is
	a core of $\Pi_{O,U,\mathit{core}_N}$ and a subset of $N$, whenever $\aprogram_{\anoverapproximation,\anunderapproximation,\mathit{core}_{\achunk}}$ is incoherent; and is an answer set of $\Pi_{O,U,\mathit{core}_N}$, otherwise.
Then, we define $\innerpropagators$ as the set of rules of
Figure~\ref{fig:trcontrolrules}.
Therefore, we consider a graph
$\mathit{FS}_\aprogram=
(V_\atoms{\aprogram},
\{\mathit{CoreOracle}\}
\setunion\innerpropagators)$
which represents Algorithm 6 in~\cite{JanotaLM15}.
Here, we need to introduce two intermediate control states: $\prelfailinstruction_{N}$ and $\mathit{Eval}$. In particular, $\prelfailinstruction_{N}$
is reached in case of inconsistency,
where $N$ is the set of literals that may be used for the
potential upcoming $\mathit{core}$ action; while $\mathit{Eval}$ is reached in case of consistency. 
From an outermost state, of the type $\mathit{Eval}$, a
new $\mathit{core}$ is started with $\newsetrule$, whenever there is a gap between over- and under-approximation; otherwise, the $\mathit{Final}$ control rule leads to the terminal state. 
$\failrule^1_{\prelaction{}}$ and $\failrule^2_{\prelaction{}}$ lead to the intermediate type of control state, $\mathit{Pre}_N$, that can either restart a
$\mathit{core}$ action with $Continue$, or continue with the $Main$ rule. Figure~\ref{fig:example-core} shows a possible path in $\mathit{FS}_\Pi$ for the program $\Pi$ of Example~\ref{ex:cautious}.

\begin{figure}[t] \footnotesize{
		$$
		\arraycolsep=3pt
		\begin{array}{llll}
		\multicolumn{4}{l}{\textrm{Return rules}}\\
				
			\mathit{Fail}^1_{pre}
		& L_{O,U,\mathit{core}_N}
		& \transitionarrow
		\mathit{Pre}_{N\setminus \{l\}}(O, U\cup\{\bar{l}\})
		& \textrm{if}\left\{
		\begin{array}{l}
			L \mbox{ is inconsistent and } \widehat{L}\cap N =\{l\}
		\end{array}
		\right.\\
		
			\mathit{Fail}^2_{pre}
		& L_{O,U,\mathit{core}_N}
		& \transitionarrow
		\mathit{Pre}_{N\setminus \widehat{L}}(O, U)
		& \textrm{if}\left\{
		\begin{array}{l}
			L \mbox{ is inconsistent and } |\widehat{L}\cap N| > 1
		\end{array}
		\right.\\
				
		\succeedrule_{\prelaction{}}
		& L_{O,U,\mathit{core}_N}
		& \transitionarrow
		\mathit{Eval}(O\cap L,U)
		& \textrm{if}\left\{
		\begin{array}{l}
			\textrm{} L \mbox{ is consistent and } L\neq\emptyset
		\end{array}
		\right.\\
		\\
		
		\multicolumn{4}{l}{\textrm{Control rules}}\\
		Main
		& \mathit{Pre}_N(O,U)
		& \transitionarrow
		\mathit{Cont}(O,U)
		& \textrm{if}\left\{
		\begin{array}{l}
			\achunk=\emptyset
		\end{array}
		\right.\\
		
		Continue
		& \mathit{Pre}_N(O,U)
		& \transitionarrow
		\emptyset_{O,U,\mathit{core}_N}
		& \textrm{if}\left\{
		\begin{array}{l}
			\textrm{} N\neq \emptyset
		\end{array}
		\right.\\
		
		\mathit{NewSet}
		& \mathit{Eval}(O,U)
		& \transitionarrow
		\emptyset_{O,U,\mathit{core}_{\overline{O}} }       
		&  \textrm{if}\left\{
		\begin{array}{l}
			O\neq U
		\end{array}
		\right.\\
		
		\mathit{Final}
		& \mathit{Eval}(O,U)
		& \transitionarrow
		\mathit{Ok} (O)       
		&  \textrm{if}\left\{
		\begin{array}{l}
			O= U
		\end{array}
		\right.\\
	\end{array}
	$$
}
\caption{The transition rules of $\mathit{in}$.}
\label{fig:trcontrolrules}
\end{figure}
\normalsize

\begin{theorem}
	Let $\Pi$ be a program, and let $O$ and $U$ be two set of atoms. Then,
	$(i)$ the only reachable terminal states are either $Cont(O,U)$ or $Ok(O)$;
	$(ii)$ if $Ok(O)$ is reachable in $FS_\Pi$, 
	then $FS_\Pi$ solves cautious reasoning;
	$(iii)$ if $Cont(O,U)$ is reachable in $FS_\Pi$,
	then $U\subseteq \mathit{cautious}(\Pi)\subseteq O$. 
\end{theorem}

\begin{figure}[t]
	\footnotesize{
		$$
		\arraycolsep=2pt
		\begin{array}{l|l}
		\begin{array}{ll}
		\multicolumn{2}{l}{  
			\Pi_{
				\{\anatom,\anotheratom,\athirdatom\},\emptyset,
				\mathit{core}_{\{\lnot a,\lnot b,\lnot c\}}}
			= \left\{
			\begin{array}{l}
			\anatom  \aspimplication \logicalnot\anotheratom\\
			\anotheratom  \aspimplication \logicalnot\anatom\\
			\athirdatom  \aspimplication \anatom\\
			\athirdatom  \aspimplication \anotheratom
			\end{array}\right\} 
			\cup
			\left\{
			\begin{array}{l} 
			\aspimplication a\\
			\aspimplication b\\
			\aspimplication c
			\end{array}\right\} }\\
		\\
		\\
		
		\multicolumn{2}{l}{ 
			\Pi_{
				\{\anatom,\anotheratom,\athirdatom\},\{c\},
				\mathit{core}_{\{\lnot a,\lnot b\}}}
			= \left\{
			\begin{array}{l}
				\anatom  \aspimplication \logicalnot\anotheratom\\
				\anotheratom  \aspimplication \logicalnot\anatom\\
				\athirdatom  \aspimplication \anatom\\
				\athirdatom  \aspimplication \anotheratom
			\end{array}\right\}
			\cup
			\left\{
			\begin{array}{l}
				\aspimplication a\\
				\aspimplication b\\
			\end{array}\right\}} \\
	\end{array}
	
	&
	
	\begin{array}{ll}
		& 		\emptyset_{\{\anatom,\anotheratom,\athirdatom\},\emptyset,\mathit{core}_{\{\lnot a,\lnot b,\lnot c\}}}
		\\
		\mathit{CoreOracle}\mbox{ :}
		&
		c\lnot c_{\{\anatom,\anotheratom,\athirdatom\},\emptyset,\mathit{core}_{\{\lnot a,\lnot b,\lnot c\}}}
		\\
		\mathit{Fail}^1_{pre}\mbox{ :}
		&
		\mathit{Pre}_{\{\lnot a,\lnot b\}}(\{a,b,c\},\{c\})
		\\
		\mathit{Continue}\mbox{ :}
		&
		\emptyset_{\{\anatom,\anotheratom,\athirdatom\},\{c\},
			\mathit{core}_{\{\lnot a,\lnot b\}}}
		\\
		\\
		\mathit{CoreOracle}\mbox{ :}
		&
		ab\lnot a\lnot b_{\{\anatom,\anotheratom,\athirdatom\},\{c\},
			\mathit{core}_{\{\lnot a,\lnot b\}}}
		\\
		\mathit{Fail}^2_{pre}\mbox{ :}
		&
		\mathit{Pre}_\emptyset(\{a,b,c\},\{c\})
		\\
		\mathit{Main}\mbox{ :}
		&
		\mathit{Cont}(\{a,b,c\},\{c\})
	\end{array}
\end{array}
$$
}
\caption{A path in $\mathit{FS}_\Pi$.}
\label{fig:example-core}
\end{figure}

Chunking and core-based methods can be combined using our methodology to abstract Algorithm 7 from~\cite{JanotaLM15}. Such a combination will be employed in the experiments.

\section{Experimental Analysis}
\label{sec:exp}
The abstract solvers reported in this paper have been used for implementing several algorithms in the ASP solver \textsc{wasp} \cite{alv15,DBLP:conf/lpnmr/AlvianoADLMR19}, resulting in the following new versions of \textsc{wasp}:
\begin{itemize}
	\item \textsc{wasp-chunk-2}, i.e., \textsc{wasp} running the algorithm based on chunking, with the size of the chunk set to 2;
	\item \textsc{wasp-chunk-20\%}, i.e., \textsc{wasp} running the algorithm based on chunking, with the size of the chunk set to the 20\% of the initial number of candidates, where the initial set of candidates is the whole set of atoms;
	\item \textsc{wasp-cb}, i.e., \textsc{wasp} running the algorithm based on cores.
	\item \textsc{wasp-cb-2}, i.e., \textsc{wasp} running the algorithm based on cores and chunking, with the size of the chunk set to 2;
	\item \textsc{wasp-cb-20\%}, i.e., \textsc{wasp} running the algorithm based on cores and chunking, with the size of the chunk set to the 20\% of the initial number of candidates.
\end{itemize}

\paragraph{Benchmark selection.}
The performance of these versions of \textsc{wasp} was measured on the benchmarks considered in~\cite{AlvianoDJMP18}.
In particular, \cite{AlvianoDJMP18} includes \textit{(i)} 
all the 193 instances from the latest ASP Competitions \cite{DBLP:journals/tplp/CalimeriIR14,DBLP:journals/ai/CalimeriGMR16,GebserMR17} involving non-ground queries;
\textit{(ii)} 115 instances of ASP Competitions classified as \textit{easy}, that is, those for which a stable model is found within 20 seconds of computation by mainstream ASP systems; and
\textit{(iii)} instances from abstract argumentation frameworks submitted to the 2nd International Competition on Computational Models of Argumentation.
In this paper, instances from \textit{(iii)} are not included since they are trivial for all tested solvers \cite{AlvianoDJMP18}.

\paragraph{Compared approaches.}
As a reference to the state of the art, we used \textsc{clasp} v. 3.3.3 \cite{GebserKS12}, which implements algorithm $\textsc{or}$ (i.e., \textit{over-approximation}), and the best performing algorithms implemented by \textsc{wasp} \cite{AlvianoDR14,AlvianoDJMP18}, namely $\textsc{or}$ (i.e., \textit{over-approximation}), $\textsc{ict}$ (i.e., \textit{under-approximation}), $\textsc{opt},$ and $\textsc{cm}$.

Algorithm $\textsc{opt}$ was presented in~\cite{AlvianoDJMP18}. The idea is as follows. Given a set of objective atoms $A$, the branching heuristic of the solver is forced to select $\neg p$ for $p \in A$, before any other unassigned literal.
In this way, the search is driven to falsify as many atoms in $A$ as possible.
When all atoms in $A$ are assigned, standard answer set search procedure is applied without further modifications to the branching heuristic.
Therefore, whenever an answer set is found, it is guaranteed to be minimal with respect to the set of objective atoms \cite{DBLP:journals/constraints/RosaGM10}.
When the current assignment to atoms in $A$ cannot be extended to an answer set, then the assignment of some atom in $A$ is flipped, and hence the procedure is repeated with a different assignment for the objective atoms.
For cautious reasoning, $A$ is initialized to the set of all candidates and updated whenever an answer set is found.

Algorithm $\textsc{cm}$ was also presented in~\cite{AlvianoDJMP18} and is based on the property that an atom is a cautious consequence of a given program if and only if the negation of the atom is an unsatisfiable core.
Hence, the algorithm searches for an answer set falsifying all candidates, with the aim of eliminating all remaining candidates at once. As soon as
no such an answer set exists, the returned unsatisfiable core is either minimized to a singleton or used to discard candidates.

Note that all tested algorithms take advantage of the incremental interface of \textsc{clasp} and \textsc{wasp}, which is based on the concept of \textit{assumptions literals}. The incremental interface allows the solver to reuse part of the computation among different calls, e.g., learned constraints and heuristic parameters.

The \textsc{dlv} solver is not considered here as its performance on cautious reasoning has been shown in earlier work to be dominated by
the other approaches considered in~\cite{AlvianoDR14}.

\paragraph{Hardware configurations and limits.}
The experiments were run on computing nodes with Intel Xeon 2.4-GHz processors and 16 GB of memory.
Time and memory limits were set to 600 seconds and 15 GB, 
respectively.

\begin{figure}[t]
	\figrule
	\begin{tikzpicture}[scale=0.85]
	\pgfkeys{%
		/pgf/number format/set thousands separator = {}}
	\begin{axis}[
	scale only axis
	, font=\normalsize
	, x label style = {at={(axis description cs:0.5,0.0)}}
	, y label style = {at={(axis description cs:0.0,0.5)}}
	, xlabel={Number of solved instances}
	, ylabel={Per-instance time limit (s)}
	, xmin=15, xmax=200
	, ymin=0, ymax=620
	, legend style={at={(0.16,0.96)},anchor=north, draw=none,fill=none}
	, legend columns=1
	, width=1\textwidth
	, height=0.4\textwidth
	, ytick={0,150,300,450,600}
	, major tick length=2pt
	]
	
	\addplot [mark size=3pt, color=green, mark=star] [unbounded coords=jump] table[col sep=semicolon, y index=1] {./qa.csv};
	\addlegendentry{\textsc{clasp}}
	\addplot [mark size=3pt, color=blue, mark=x] [unbounded coords=jump] table[col sep=semicolon, y index=7] {./qa.csv};
	\addlegendentry{\textsc{wasp-or}}
	\addplot [mark size=3pt, color=blue, mark=square] [unbounded coords=jump] table[col sep=semicolon, y index=8] {./qa.csv};
	\addlegendentry{\textsc{wasp-ict}}	
	\addplot [mark size=3pt, color=blue, mark=square*] [unbounded coords=jump] table[col sep=semicolon, y index=9] {./qa.csv};
	\addlegendentry{\textsc{wasp-cm}}	
	\addplot [mark size=3pt, color=blue, mark=triangle*] [unbounded coords=jump] table[col sep=semicolon, y index=10] {./qa.csv};
	\addlegendentry{\textsc{wasp-opt}}	 
	\addplot [mark size=3pt, color=black, mark=o] [unbounded coords=jump] table[col sep=semicolon, y index=5] {./qa.csv};
	\addlegendentry{\textsc{wasp-chunk-20\%}}
	\addplot [mark size=3pt, color=black, mark=*] [unbounded coords=jump] table[col sep=semicolon, y index=6] {./qa.csv};
	\addlegendentry{\textsc{wasp-chunk-2}}
	\addplot [mark size=3pt, color=red, mark=triangle] [unbounded coords=jump] table[col sep=semicolon, y index=2] {./qa.csv};
	\addlegendentry{\textsc{wasp-cb}}
	\addplot [mark size=3pt, color=red, mark=o] [unbounded coords=jump] table[col sep=semicolon, y index=3] {./qa.csv};
	\addlegendentry{\textsc{wasp-cb-20\%}}
	\addplot [mark size=3pt, color=red, mark=*] [unbounded coords=jump] table[col sep=semicolon, y index=4] {./qa.csv};
	\addlegendentry{\textsc{wasp-cb-2}}	
	\end{axis}
	\end{tikzpicture}
	
	\caption{Benchmark \textit{(i)}: Performance comparison on non-ground queries in ASP Competitions.}\label{fig:cactusi}
	\figrule
\end{figure}
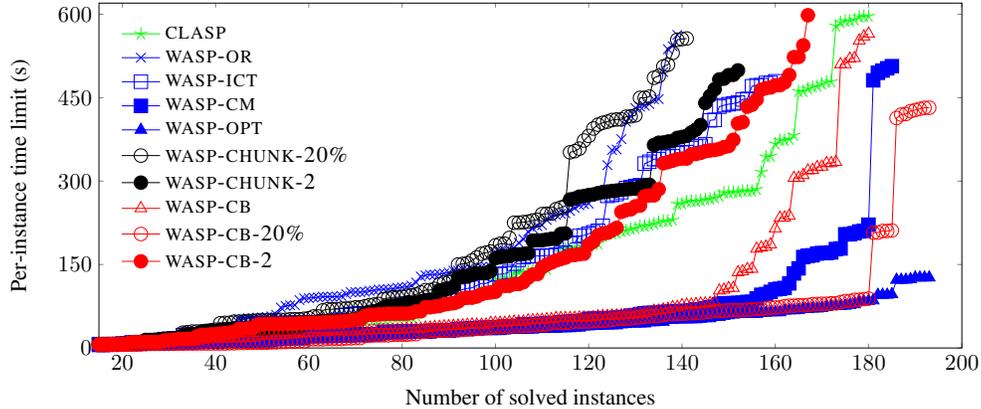

\subsection{Results}
Concerning benchmark \textit{(i)}, results are shown in the cactus plot of Figure~\ref{fig:cactusi}, where for each algorithm the number of solved instances in a given time is reported, producing an aggregated view of its overall performance.
As a first observation, \textsc{wasp} cannot reach the performance of \textsc{clasp} on the execution of algorithm \textsc{or}, and indeed \textsc{clasp} solved 41 instances more than \textsc{wasp-or}.
However, such a huge gap is completely filled by \textsc{wasp-cb-20\%}, which actually solves 13 instances more than \clasp.
Indeed, \textsc{wasp-cb-20\%} is able to solve all instances with an average running time of 56 seconds, and is comparable to the best performing algorithm, namely \textsc{wasp-opt}, which solves all instances with an average running time of 36 seconds.
Notably, even a small size of the chunk may have a huge impact on the performance of the algorithms. Indeed, \textsc{wasp-cb} outperforms \textsc{wasp-cb-2}, solving 13 instances more.
Finally, we observe that \textsc{wasp-chunk-20\%} and \textsc{wasp-chunk-2} are not competitive with algorithms based on cores.

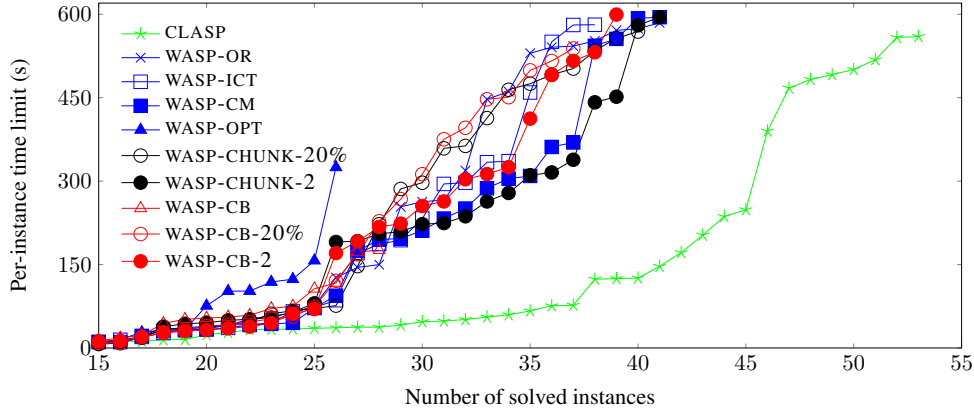
\begin{figure}[t]
	\figrule
	\begin{tikzpicture}[scale=0.85]
	\pgfkeys{%
		/pgf/number format/set thousands separator = {}}
	\begin{axis}[
	scale only axis
	, font=\normalsize
	, x label style = {at={(axis description cs:0.5,0.0)}}
	, y label style = {at={(axis description cs:0.0,0.5)}}
	, xlabel={Number of solved instances}
	, ylabel={Per-instance time limit (s)}
	, xmin=15, xmax=55
	, ymin=0, ymax=620
	, legend style={at={(0.16,0.96)},anchor=north, draw=none,fill=none}
	, legend columns=1
	, width=1\textwidth
	, height=0.4\textwidth
	, ytick={0,150,300,450,600}
	, major tick length=2pt
	]
	
	\addplot [mark size=3pt, color=green, mark=star] [unbounded coords=jump] table[col sep=semicolon, y index=1] {./easy.csv};
	\addlegendentry{\textsc{clasp}}
	\addplot [mark size=3pt, color=blue, mark=x] [unbounded coords=jump] table[col sep=semicolon, y index=7] {./easy.csv};
	\addlegendentry{\textsc{wasp-or}}
	\addplot [mark size=3pt, color=blue, mark=square] [unbounded coords=jump] table[col sep=semicolon, y index=8] {./easy.csv};
	\addlegendentry{\textsc{wasp-ict}}	
	\addplot [mark size=3pt, color=blue, mark=square*] [unbounded coords=jump] table[col sep=semicolon, y index=9] {./easy.csv};
	\addlegendentry{\textsc{wasp-cm}}	
	\addplot [mark size=3pt, color=blue, mark=triangle*] [unbounded coords=jump] table[col sep=semicolon, y index=10] {./easy.csv};
	\addlegendentry{\textsc{wasp-opt}}	 
	\addplot [mark size=3pt, color=black, mark=o] [unbounded coords=jump] table[col sep=semicolon, y index=5] {./easy.csv};
	\addlegendentry{\textsc{wasp-chunk-20\%}}
	\addplot [mark size=3pt, color=black, mark=*] [unbounded coords=jump] table[col sep=semicolon, y index=6] {./easy.csv};
	\addlegendentry{\textsc{wasp-chunk-2}}
	\addplot [mark size=3pt, color=red, mark=triangle] [unbounded coords=jump] table[col sep=semicolon, y index=2] {./easy.csv};
	\addlegendentry{\textsc{wasp-cb}}
	\addplot [mark size=3pt, color=red, mark=o] [unbounded coords=jump] table[col sep=semicolon, y index=3] {./easy.csv};
	\addlegendentry{\textsc{wasp-cb-20\%}}
	\addplot [mark size=3pt, color=red, mark=*] [unbounded coords=jump] table[col sep=semicolon, y index=4] {./easy.csv};
	\addlegendentry{\textsc{wasp-cb-2}}	
	\end{axis}
	\end{tikzpicture}
	
	\caption{Benchmark \textit{(ii)}: Performance comparison on computation of cautious consequences for \emph{easy} instances of ASP Competitions.}\label{fig:cactusii}
	\figrule
\end{figure}

Concerning benchmark \textit{(ii)}, results are shown in the cactus plot of Figure~\ref{fig:cactusii}.
It is possible to observe that \textsc{clasp} is the best performing solver on this benchmark, solving 53 instances overall.
If we focus on \textsc{wasp}, the best performance is obtained by \textsc{wasp-chunk-2}, \textsc{wasp-or}, \textsc{wasp-cm}, and \textsc{wasp-chunk-20\%} which are able to solve 41, 41, 41, and 40 instances, respectively.
Moreover, \textsc{wasp-cb}  cannot reach the same performance on this benchmark, solving only 25 instances.
We observe that the poor performance depends on the first calls to the oracle, since they are expensive in terms of solving time.
This negative effect is mitigated by chunking since \textsc{wasp-cb-20\%} and \textsc{wasp-cb-2} solve 37 and 39 instances, respectively.

Finally, detailed results of benchmarks \textit{(i)} and \textit{(ii)} are shown in Table~\ref{tab:queryanswering}, where we report the 5 algorithms solving the largest number of instances. In particular, for each algorithm we report the number of solved instances and the cumulative solving time (for each timeout we added 600 seconds).
We also observe that \textsc{wasp-cb-20\%} is comparable with \textsc{clasp} solving only 3 instances  less.

\begin{table}[b!]
	\caption{
		Numbers of solved instances and cumulative running time (in seconds; each timeout adds 600 seconds) on instances from benchmarks \textit{(i)} and \textit{(ii)}.
	}
	\label{tab:queryanswering}
	\centering
	\tabcolsep=0.040cm
	\begin{tabular}{lrrrrrrrrrrrrrrrr}
		\toprule
		&  && \multicolumn{2}{c}{\textbf{\textsc{clasp}}}  && \multicolumn{2}{c}{\textbf{\textsc{wasp-cm}}} && \multicolumn{2}{c}{\textbf{\textsc{wasp-opt}}} && \multicolumn{2}{c}{\textbf{\textsc{wasp-cb}}}  && \multicolumn{2}{c}{\textbf{\textsc{wasp-cb-20\%}}} \\
		\cmidrule{4-5}		\cmidrule{7-8} 		\cmidrule{10-11} 		\cmidrule{13-14} 		\cmidrule{16-17}
		\textbf{Benchmark} & \textbf{\#} && \textbf{sol.} & \textbf{sum t} && \textbf{sol.} & \textbf{sum t} && \textbf{sol.} & \textbf{sum t} && \textbf{sol.} & \textbf{sum t} && \textbf{sol.} & \textbf{sum t}\\	
		\cmidrule{1-17}
		CQA-Q3 & 40 && 40 & 4354 && 40 & 1313 && 40 & 1276 && 40 & 1291 && 40 & 1303\\
		CQA-Q6 & 40 && 40 & 8505 && 40 & 2149 && 40 & 1956 && 40 & 3544 && 40 & 1849\\
		CQA-Q7 & 40 && 40 & 8929 && 40 & 1741 && 40 & 1681 && 40 & 1735 && 40 & 1724\\
		MCSQ & 73 && 60 & 12701 && 65 & 11007 && 73 & 1995 && 60 & 15757 && 73 & 5924\\
		\cmidrule{1-17}
		GracefulGraphs & 1 && 1 & 51 && 1 & 45 && 1 & 32 && 1 & 44 && 1 & 57\\
		GraphCol & 1 && 0 & 600 && 0 & 600 && 0 & 600 && 0 & 600 && 0 & 600\\
		IncrSched & 6 && 5 & 857 && 2 & 2692 && 1 & 3016 && 1 & 3006 && 1 & 3004\\
		KnightTour & 2 && 2 & 62 && 0 & 1200 && 0 & 1200 && 0 & 1200 && 0 & 1200\\
		Labyrinth & 32 && 6 & 18377 && 0 & 19200 && 0 & 19200 && 0 & 19200 && 1 & 18912\\
		NoMystery & 2 && 1 & 1091 && 1 & 694 && 0 & 1200 && 1 & 706 && 1 & 721\\
		PPM & 15 && 15 & 264 && 15 & 81 && 15 & 76 && 15 & 113 && 15 & 76\\
		QualSpatReas & 18 && 18 & 1019 && 17 & 4537 && 7 & 7406 && 7 & 7083 && 14 & 5707\\
		Sokoban & 36 && 3 & 20529 && 3 & 20665 && 1 & 21102 && 1 & 21023 && 2 & 20918\\
		VisitAll & 2 && 2 & 80 && 2 & 408 && 1 & 757 && 2 & 348 && 2 & 396\\
		\cmidrule{1-17}
		\textbf{Total} & \textbf{308} && \textbf{233} & \textbf{78584} && \textbf{226} & \textbf{66931} && \textbf{219} & \textbf{62097} && \textbf{208} & \textbf{76252} && \textbf{230} & \textbf{62993}\\
		\bottomrule
	\end{tabular}
\end{table}

\section{Related Work}
\label{sec:related}
Abstract solvers methodology for describing solving procedures have been introduced for the {\dpll} procedure with learning of SAT solving and for certain extensions implemented in SMT solvers \cite{nie06}. In ASP, \citeN{lier08} introduced and compared the abstract solvers for {\sc smodels} and {\cmodels} on non-disjunctive programs, then in~\cite{lier10} the framework has been extended by introducing transition rules that capture backjumping and learning techniques.
\citeN{lier11} presented a unifying perspective based on completion of solvers for non-disjunctive answer set solving.
\citeN{blm14} presented abstract solvers for disjunctive answer set solvers {\cmodels}, {\gnt} and {\dlv} implementing plain backtracking, and~\citeN{lie14} defined abstract frameworks for Constraint ASP solvers.

All these papers describe ASP procedures for computing (one) stable models in abstract solvers methodology.
In our paper we have, instead, focused on the description of ASP procedures for cautious reasoning tasks, possibly employing some of the solutions presented in related papers as ASP oracle calls.
Our paper significantly extends the short technical communication \cite{bro15b} by $(i)$ designing more advanced solving techniques, like chunking and core-based algorithms, that lead to new solving solutions, $(ii)$ implementing and testing such new solutions, $(iii)$ adding further examples and a detailed related work, and $(iv)$ formally stating a strong analogy between backbones computation in SAT and cautious reasoning in ASP.

As far as the application of abstract solvers methodology outside ASP is concerned, the first application has been already mentioned and is related to the seminal paper~\cite{nie06}, where SMT problems with certain logics via a lazy approach~\cite{Sebastiani07} are considered. Then, abstract
solvers have been presented for the satisfiability of Quantified Boolean Formulas by~\citeN{bro15}, and for solving certain reasoning taks in Abstract Argumentation under preferred semantics~\cite{BrocheninLMWW18}. Finally, in another number of papers, starting from a developed concept of modularity in answer set solving~\cite{LierlerT13}, abstract modeling of solvers for multi-logic systems are presented~\cite{LierlerT14,LierlerT15,LierlerT16}.

Another added, general, value of our paper is in its practical part, i.e., an implementation of new solutions designed through abstract solvers. In fact, while nowadays abstract solvers methodology has been widely used, often in the mentioned papers the presented results have rarely led to implementations, with the exception of~\cite{nie06}, where the related {\sc Barcelogic} implementation won the SMT Competition 2005 on same logics, and~\cite{lier10}, where a proposed combination of {\sc smodels} and {\sc cmodels} techniques has been implemented in the solver {\sc sup}, that reached positive results at the ASP Competition 2011 and, more recently,~\cite{BrocheninLMWW18}, where the new designed solution, obtained as a modification of the {\sc cegartix} solver, performed often better than the basic {\sc cergatix} solver on preferred semantics, that was among the best solvers in the first ICCMA competition.

Finally, very recently improved algorithms for computing cautious consequences of ASP programs have been presented in~\cite{AlvianoDJMP18}: such algorithms could be also modeled through abstract solvers and combined with the ones presented in this paper.

\section{Conclusion}
\label{sec:concl}
In this paper we modeled through abstract solvers advanced techniques for solving cautious reasoning tasks in ASP. Such advanced techniques have been borrowed from the computation of backbones of propositional formulas. We have then designed new solving procedures, and implemented them in {\sc wasp}, that already included algorithms of~\cite{AlvianoDR14,AlvianoDJMP18}. Experiments on devoted benchmarks have shown positive results for the new proposed solutions. At the same time, our work has formally stated, through an uniform treatment, a strong analogy among the algorithms for computing backbones of propositional formulas and those for computing cautious consequences of ASP programs.
Finally, we remark that algorithms presented in this paper are independent with respect to the underlying solving strategies, and can be complemented with existing heuristics and optimization techniques~\cite{DBLP:conf/jelia/GiunchigliaMT02,DBLP:conf/cp/GiunchigliaMT03,giu08}.

\bibliographystyle{acmtrans}
\bibliography{shared}    

\clearpage

 \appendix
\section{Proofs}
\subsection{Correctness of the Oracle}
\begin{paragraph}{Definitions.}
	For a program $\aprogram$ and a type of model
	$\atypeofmodel\in\{\classic,\stable\}$, we say that $\amodel$ is a
	\emph{$\atypeofmodel$-model of $\aprogram$} when either $\atypeofmodel$ is
	$\stable$ and $\amodel$ is a stable model of $\aprogram$ or $\atypeofmodel$ is
	$\classic$ and $\amodel$ is a classical model of $\aprogram$.
	We define
	$\asetofliterals_{\classic}=\atoms{\aprogram}$
	and $\asetofliterals_{\stable}=\atoms{\aprogram}$.
	Also $\anaim_{\classic}=\backboneof{\aprogram}$
	and $\anaim_{\stable}=\cautiousof{\aprogram}$.
	We say that $(\aprogram,\atypeofmodel,S,\agraph)$ is a \emph{suitable quadruple}
	when $\aprogram$ is a program,
	$\atypeofmodel\in\{\classic,\stable\}$,
	$S\subseteq\{\underpropagators,\overpropagators,\chunkpropagators\}$, and
	$\agraph=(\statesof{\atoms{\aprogram}},
	\{\mathit{Oracle}\}
	\setunion\bigcup_{x\in S}x
	)$.
\end{paragraph}

\begin{lemma}\label{lemma:path}
	Let $(\aprogram,\atypeofmodel,S,\agraph)$ be a suitable quadruple, and
	let
	$\corestate{\astringofliterals}{\anoverapproximation}{\anunderapproximation}{\anaction}{}$
	be a reachable state from
	$\corestate{\emptyset}
	{\asetofliterals_\atypeofmodel}{\emptyset}
	{B}{\logicalfalse}$ in $\agraph$, where $B\in\{\mathit{over},\mathit{under}_\emptyset,\initaction\}$.
	There is a path in $\agraph$ from
	$\corestate{\emptyset}{\anoverapproximation}{\anunderapproximation}{A}{}$
	to
	$\corestate{\astringofliterals}{\anoverapproximation}{\anunderapproximation}{\anaction}{}$
	that does not contain any control state.
\end{lemma}
\begin{proof}
	Let
	$\corestate{\astringofliterals}{\anoverapproximation}{\anunderapproximation}{\anaction}{}$
	be a state reachable from
	$\corestate{\emptyset}
	{\asetofliterals_\atypeofmodel}{\emptyset}
	{B}{\logicalfalse}$ in $G$, where $B\in\{\mathit{over},\mathit{under}_\emptyset,\initaction\}$.
	Assume it is reachable without going through any control state; in this case
	$\anaction = B$, $\anunderapproximation=\emptyset$ and
	$\anoverapproximation=\asetofliterals_\atypeofmodel$ as the $\mathit{Oracle}$ rule does not modify these.
	Otherwise a path $H$ leading to
	$\corestate{\astringofliterals}{\anoverapproximation}{\anunderapproximation}{\anaction}{}$
	goes through some control state; and after the last control state in this path,
	a rule among $\{\underapproxrule,$ $\overapproxrule,$ $\chunkapproxrule\}$ has been
	applied, which involves that the state occurring right after applying this rule
	was $\corestate{\emptyset}{\anoverapproximation'}{\anunderapproximation'}{\anaction'}{}$
	for some $\anoverapproximation'$, $\anunderapproximation'$ and $\anaction'$.
	The $\mathit{Oracle}$ rule does not modify these components of oracle states, and
	additionally, by the choice of the last control state in $H$ as the
	predecessor of
	$\corestate{\emptyset}{\anoverapproximation'}{\anunderapproximation'}{\anaction'}{}$,
	there is no control state in the part of $H$ from
	$\corestate{\emptyset}{\anoverapproximation'}{\anunderapproximation'}{\anaction'}{}$
	to $\corestate{\astringofliterals}{\anoverapproximation}{\anunderapproximation}{\anaction}{}$.
	So necessarily $\anoverapproximation'=\anoverapproximation$,
	$\anunderapproximation'=\anunderapproximation$ and $\anaction'=\anaction$.
	Hence, in any case there is a path from
	$\corestate{\emptyset}{\anoverapproximation}{\anunderapproximation}{\anaction}{}$
	to $\corestate{\astringofliterals}{\anoverapproximation}{\anunderapproximation}{\anaction}{}$
	that does not contain any control state.
\end{proof}

\begin{lemma}\label{lemma:return}
	Let $(\aprogram,\atypeofmodel,S,\agraph)$ be a suitable quadruple, and
	let
	$\corestate{\astringofliterals}{\anoverapproximation}{\anunderapproximation}{\anaction}{}$
	be a reachable state from
	$\corestate{\emptyset}
	{\asetofliterals_\atypeofmodel}{\emptyset}
	{B}{\logicalfalse}$ in $\agraph$, where $B\in\{\mathit{over},\mathit{under}_\emptyset,\initaction\}$.
	If the rule
	$\failrule_\anaction$ applies to
	$\corestate{\astringofliterals}{\anoverapproximation}{\anunderapproximation}{\anaction}{}$
	in $\agraph$, then
	$\aprogram_{\anoverapproximation,\anunderapproximation,\anaction}$ has no
	$\atypeofmodel$-model; and, if the rule $\succeedrule$ applies, then
	$\astringofliterals$ is a $\atypeofmodel$-model of
	$\aprogram_{\anoverapproximation,\anunderapproximation,\anaction}$.
\end{lemma}
\begin{proof}
	By Lemma \ref{lemma:path}, there is a path from
	$\corestate{\emptyset}{\anoverapproximation}{\anunderapproximation}{\anaction}{}$
	to
	$\corestate{\astringofliterals}{\anoverapproximation}{\anunderapproximation}{\anaction}{}$
	that does not contain any control state. Hence, this path is justified
	exclusively by the $\mathit{Oracle}$ rule.
	
	First, assume that $\atypeofmodel=\classic$.
	Applying the results from \citeN{nie06}, the lemma holds in this case. If
	the rule $\failrule_\anaction$ applies to
	$\corestate{\astringofliterals}{\anoverapproximation}{\anunderapproximation}{\anaction}{}$
	in $\agraph$, then
	$\aprogram_{\anoverapproximation,\anunderapproximation,\anaction}$ has no
	classical model, and if the rule $\succeedrule$ applies then
	$\astringofliterals$ is a classical model of
	$\aprogram_{\anoverapproximation,\anunderapproximation,\anaction}$. 
	
	Second, assume that $\atypeofmodel=\stable$. 
	Then, by the results of \citeN{LierlerT14} the
	Lemma also holds in this case. Indeed, if the
	rule $\failrule_\anaction$ applies to
	$\corestate{\astringofliterals}{\anoverapproximation}{\anunderapproximation}{\anaction}{}$
	in $\agraph$, then
	$\aprogram_{\anoverapproximation,\anunderapproximation,\anaction}$ has no
	stable model; and if the rule $\succeedrule$ applies, then
	$\astringofliterals$ is a stable model of
	$\aprogram_{\anoverapproximation,\anunderapproximation,\anaction}$. 
\end{proof}

\subsection{Correctness of the Structure}
\begin{lemma}\label{lemma:struct1}
	Let $(\aprogram,\atypeofmodel,S,\agraph)$ be a suitable quadruple, and
	if a state
	$\corestate{\astringofliterals}{\anoverapproximation}{\anunderapproximation}{\anaction}{}$
	or $\controlstate{\continueinstruction}{\astringofliterals}{\anoverapproximation}{\anunderapproximation}{\anaction}{}$
	is reachable from
	$\corestate{\emptyset}
	{\asetofliterals_\atypeofmodel}{\emptyset}
	{B}{\logicalfalse}$ in $\agraph$, where $B\in\{\mathit{over},\mathit{under}_\emptyset,\initaction\}$,
	then
	$\anunderapproximation\subseteq\anaim_\atypeofmodel\subseteq\anoverapproximation$.
\end{lemma}
\begin{proof}
	We prove this lemma by induction on the path leading from 
	$\corestate{\emptyset}
	{\asetofliterals_\atypeofmodel}{\emptyset}
	{B}{\logicalfalse}$ to
	$\corestate{\astringofliterals}{\anoverapproximation}{\anunderapproximation}{\anaction}{}$
	or
	$\controlstate{\continueinstruction}{\astringofliterals}{\anoverapproximation}{\anunderapproximation}{\anaction}{}$.
	So as to initialize this induction, we simply note that
	$\corestate{\emptyset}
	{\asetofliterals_\atypeofmodel}{\emptyset}
	{B}{\logicalfalse}$
	is such that
	$\emptyset\subseteq\anaim_\atypeofmodel\subseteq\asetofliterals_\atypeofmodel$.
	Now, assume that a state
	is reachable from
	$\corestate{\emptyset}
	{\asetofliterals_\atypeofmodel}{\emptyset}
	{B}{\logicalfalse}$ in $\agraph$ and that for any state on
	the path the lemma holds, in particular on its predecessor. 
	We are going to prove that for this state the lemma holds.
	
	First case: assume that the state is a core state
	$\corestate{\astringofliterals}{\anoverapproximation}{\anunderapproximation}{\anaction}{}$.
	If its predecessor is a core state, then the predecessor is
	$\corestate{\astringofliterals'}{\anoverapproximation}{\anunderapproximation}{\anaction}{}$
	for some $\astringofliterals'$, since the $\mathit{Oracle}$ rule does not modify these $\anoverapproximation$,
	$\anunderapproximation$ and $\anaction$. 
	By the induction hypothesis, the lemma holds.
	If its predecessor is a control state then note that the control rules that may
	link this predecessor to
	$\corestate{\astringofliterals}{\anoverapproximation}{\anunderapproximation}{\anaction}{}$
	are $\overapproxrule$, $\underapproxrule$ and $\chunkapproxrule$, of which none
	modifies the over-approximation and under-approximation; hence, the predecessor is
	$\controlstate{\continueinstruction}{}{\anoverapproximation}{\anunderapproximation}{}{}$
	and by the induction hypothesis the lemma holds.
	
	Second case: when the state is a control state.
	Then, its predecessor is a core state
	$\corestate{\astringofliterals}{\anoverapproximation}{\anunderapproximation}{\anaction}{}$.
	By the induction hypothesis,
	$\anunderapproximation\subseteq\anaim_\atypeofmodel\subseteq\anoverapproximation$.
	The rule applied is a return rule.
	\begin{itemize}
		\item If the rule is $\mathit{Terminal}$, then the state is
		$\controlstate{\continueinstruction}
		{\astringofliterals}
		{\anoverapproximation\setintersection\astringofliterals}
		{\anunderapproximation}
		{\anaction}{}$.
		By Lemma \ref{lemma:return}, $\astringofliterals$ is a
		$\atypeofmodel$-model of
		$\aprogram_{\anoverapproximation,\anunderapproximation,\anaction}$. So no
		element of $\asetofliterals_\atypeofmodel\setminus\astringofliterals$
		belongs to $\anaim_\atypeofmodel$, and no element of
		$\astringofliterals$ can be part of $\anaim_\atypeofmodel$.
		Hence, $\anunderapproximation\subseteq
		\anaim_\atypeofmodel\subseteq
		\anoverapproximation\setintersection\astringofliterals$.
		\item If the rule is $\failrule_{\underapproxaction{}}$, then the state is
		$\controlstate{\continueinstruction}
		{\astringofliterals}
		{\anoverapproximation}
		{\anunderapproximation\setunion\{a\}}
		{\underapproxaction{\{\aliteral\}}}{}$.
		By Lemma \ref{lemma:return},
		$\aprogram_{\anoverapproximation,\anunderapproximation,\anaction}$ has
		no $\atypeofmodel$-model. So no $\atypeofmodel$-model of
		$\aprogram$ satisfies $a$. So $a$
		belongs to $\anaim_\atypeofmodel$.
		Hence, $\anunderapproximation\setunion\{a\}\subseteq
		\anaim_\atypeofmodel\subseteq
		\anoverapproximation$.
	\end{itemize}
	In all cases the lemma holds, which ends the proof by induction.
\end{proof}

\begin{lemma}\label{lemma:struct2}
	Let $(\aprogram,\atypeofmodel,S,\agraph)$ be a suitable quadruple, and
	let
	$\corestate{\astringofliterals}{\anoverapproximation}{\anunderapproximation}{\anaction}{}$
	be a reachable state from
	$\corestate{\emptyset}
	{\asetofliterals_\atypeofmodel}{\emptyset}
	{B}{\logicalfalse}$ in $\agraph$, where $B\in\{\mathit{over},\mathit{under}_\emptyset,\initaction\}$.
	If $\failrule_{\overapproxaction}$ applies to
	$\corestate{\astringofliterals}{\anoverapproximation}{\anunderapproximation}{\anaction}{}$,
	then $\anaim_\atypeofmodel=\anoverapproximation$.
\end{lemma}
\begin{proof}
	Assume that 
	$\failrule_{\overapproxaction}$ applies to some state
	$\corestate{\astringofliterals}{\anoverapproximation}{\anunderapproximation}{\anaction}{}$
	reachable from
	$\corestate{\emptyset}
	{\asetofliterals_\atypeofmodel}{\emptyset}
	{B}{\logicalfalse}$.
	Then $\anaction=\overapproxaction$.
	The path has to go through at least one control state so that
	$\anaction\neq B$, and hence the rule $\succeedrule$ has to have been
	applied; so $\aprogram$ has at least one $\atypeofmodel$-model and
	$\anaim_\atypeofmodel$ is well defined.
	Also, by Lemma \ref{lemma:return},
	$\aprogram_{\anoverapproximation,\anunderapproximation,\overapproxaction}$ has
	no $\atypeofmodel$-model. In other words,
	$\aprogram
	\setunion\{\aspimplication\anoverapproximation\}$
	has no $\atypeofmodel$-model.
	As the constraint added to $\aprogram$ is monotonic,
	$\aprogram$ has no $\atypeofmodel$-model satisfying
	$\aspimplication\anoverapproximation$. In other words, all the
	$\atypeofmodel$-models of $\aprogram$ satisfy $\anoverapproximation$,
	so $\anoverapproximation\subseteq\anaim_\atypeofmodel$.
	Since, by Lemma \ref{lemma:struct1},
	$\anaim_\atypeofmodel\subseteq\anoverapproximation$, also
	$\anaim_\atypeofmodel=\anoverapproximation$.
\end{proof}

\begin{lemma}\label{lemma:struct3}
	Let $(\aprogram,\atypeofmodel,S,\agraph)$ be a suitable quadruple, and
	let
	$\corestate{\astringofliterals}{\anoverapproximation}{\anunderapproximation}{\anaction}{}$
	be a reachable state from
	$\corestate{\emptyset}
	{\asetofliterals_\atypeofmodel}{\emptyset}
	{B}{\logicalfalse}$ in $\agraph$, where $B\in\{\mathit{over},\mathit{under}_\emptyset,\initaction\}$.
	If there is a transition in $\agraph$ from
	$\corestate{\astringofliterals}{\anoverapproximation}{\anunderapproximation}{\anaction}{}$
	to
	$\controlstate{\continueinstruction}{\astringofliterals}{\anoverapproximation'}{\anunderapproximation'}{\anaction}{}$
	and $\anaction\neq B$,
	then
	$\anoverapproximation'\setminus\anunderapproximation'\subset\anoverapproximation\setminus\anunderapproximation$.
\end{lemma}
\begin{proof}
	Assume that there is a transition in $\agraph$ from
	$\corestate{\astringofliterals}{\anoverapproximation}{\anunderapproximation}{\anaction}{}$
	to
	$\controlstate{\continueinstruction}{\astringofliterals}{\anoverapproximation}{\anunderapproximation}{\anaction}{}$
	and $\anaction\neq B$.
	
	If this transition is justified by 
	$\failrule_{\chunkapproxaction{}}$ or $\failrule_{\underapproxaction{}}$,
	then $\anaction$ is $\chunkapproxaction{\achunk}$
	or $\underapproxaction{\achunk}$ for some $\achunk$.
	Also $\anoverapproximation'=\anoverapproximation$ and
	$\anunderapproximation'=\anunderapproximation\setunion\achunk$, so
	$\anoverapproximation'\setminus\anunderapproximation'
	\subseteq(\anoverapproximation\setminus\anunderapproximation)\setminus\achunk$.
	The last control rule applied was necessarily $\chunkapproxrule$, so that
	$\achunk\subseteq\anoverapproximation\setminus\anunderapproximation$ and
	$\achunk\neq\emptyset$. Then
	$(\anoverapproximation\setminus\anunderapproximation)\setminus\achunk\subset\anoverapproximation\setminus\anunderapproximation$,
	so
	$\anoverapproximation'\setminus\anunderapproximation'\subset\anoverapproximation\setminus\anunderapproximation$.
	
	If this transition is justified by $\succeedrule$, we first prove that
	$\anoverapproximation\setintersection\astringofliterals\neq\anoverapproximation$
	and
	$\anunderapproximation\subseteq\astringofliterals$.
	First, assume $\anaction=\overapproxaction$.
	Then, by Lemma \ref{lemma:return}, $\astringofliterals$ is a
	$\atypeofmodel$-model of
	$\aprogram_{\anoverapproximation,\anunderapproximation,\overapproxaction}=
	\aprogram
	\setunion\{\aspimplication\anoverapproximation\}$.
	Therefore, $\astringofliterals$ is a $\atypeofmodel$-model of $\aprogram$ and a
	classical model of $\{\aspimplication\anoverapproximation\}$.
	Since it is a $\atypeofmodel$-model of $\aprogram$ and
	$\anunderapproximation\subseteq\anaim_\atypeofmodel$, by definition
	of $\anaim_\atypeofmodel$ also $\anunderapproximation\subseteq\astringofliterals$.
	Since $\astringofliterals$ is a classical model of
	$\{\aspimplication\anoverapproximation\}$, also
	$\opposite{\anoverapproximation}\setintersection\astringofliterals\neq\emptyset$.
	Hence,
	$\anoverapproximation\setintersection\astringofliterals\neq\anoverapproximation$.
	Now, assume $\anaction=\chunkapproxaction{\achunk}$.
	The last control rule applied was necessarily $\chunkapproxrule$, so that
	$\achunk\subseteq\anoverapproximation\setminus\anunderapproximation$
	and hence $\achunk\subseteq\anoverapproximation$.
	Also, by Lemma \ref{lemma:return}, $\astringofliterals$ is a
	$\atypeofmodel$-model of
	$\aprogram_{\anoverapproximation,\anunderapproximation,\chunkapproxaction{\achunk}}=
	\aprogram\setunion\{\aspimplication\achunk\}$,
	so $\astringofliterals$ is a $\atypeofmodel$-model of $\aprogram$ and a
	classical model of $\{\aspimplication\achunk\}$, and
	$\opposite{\achunk}\setintersection\astringofliterals\neq\emptyset$.
	Since it is a $\atypeofmodel$-model of $\aprogram$ and
	$\anunderapproximation\subseteq\anaim_\atypeofmodel$, by definition
	of $\anaim_\atypeofmodel$ also $\anunderapproximation\subseteq\astringofliterals$.
	Since $\astringofliterals$ is a classical model of $\{\aspimplication\achunk\}$,
	also
	$\opposite{\anoverapproximation}\setintersection\astringofliterals\neq\emptyset$, and hence
	$\anoverapproximation\setintersection\astringofliterals\neq\anoverapproximation$.
	The proof in the case of $\underapproxaction{\achunk}$ is identical to the case
	of $\chunkapproxaction{\achunk}$.
	So in any case $\anoverapproximation\setintersection\astringofliterals\neq\anoverapproximation$
	and $\anunderapproximation\subseteq\astringofliterals$. So
	$\anoverapproximation'\setminus\anunderapproximation'=
	(\anoverapproximation\setintersection\astringofliterals)\setminus\anunderapproximation$
	is a strict subset of $\anoverapproximation\setminus\anunderapproximation$.
\end{proof}

\subsection{Finiteness and Lack of Reachable Cycles}
\begin{lemma}\label{lemma:finite}	
Let $\aprogram$ be a program, and let $S\subseteq\{\underpropagators,\overpropagators,\chunkpropagators\}$.
Then, the graph $(\statesof{\atoms{\aprogram}}, \{\mathit{Oracle}\} \setunion\bigcup_{x\in S}x)$ is finite.
\end{lemma}

\begin{proof}
	Any core state relative to $\atoms{\aprogram}$ is made of a record relative to
	$\atoms{\aprogram}$, two sets of literals relative to $\atoms{\aprogram}$, and
	one action relative to $\atoms{\aprogram}$. The set
	$\literalsof{\atoms{\aprogram}}$ of literals relative to $\atoms{\aprogram}$ is
	finite, and so is its powerset; hence there is only a finite amount of
	possibilities for the two sets of literals relative to $\atoms{\aprogram}$.
	Also, since an action can only be $\overapproxaction$,
	$\chunkapproxaction{\asetofliterals}$, or $\underapproxaction{\asetofliterals}$
	for $\asetofliterals$ a set of literals relative to $\atoms{\aprogram}$, there
	is only a finite amount of possible actions. Finally, since the set of literals
	relative to $\atoms{\aprogram}$ is finite, and so is its powerset; so there are
	only a finite amount of possible records relative to $\atoms{\aprogram}$ since
	repetitions are not allowed in records. So there is a only finite amount of core
	states relative to $\statesof{\atoms{\aprogram}}$. Since the other types of states are
	only made of a portion of what makes a core state, there is also a finite amount
	of them. As a consequence, $\statesof{\atoms{\aprogram}}$ is finite, and
	hence the graph
	$(\statesof{\atoms{\aprogram}},
	\{\mathit{Oracle}\}
	\setunion\bigcup_{x\in S}x
	)$
	is finite.
\end{proof}

\begin{lemma}\label{lemma:acyclic}
	Let $(\aprogram,\atypeofmodel,S,\agraph)$ be a suitable quadruple.
	Then, there is no cycle in $\agraph$ reachable from the initial state
	$\corestate{\emptyset}
	{\asetofliterals_\atypeofmodel}{\emptyset}
	{B}{\logicalfalse}$, where $B\in\{\mathit{over},\mathit{under}_\emptyset,\initaction\}$.
\end{lemma}
\begin{proof}
	We are going to define a partial order on $\statesof{\atoms{\aprogram}}$.
	
	First, we define an order on records as follows.
	For any record $\astringofliterals$, we consider the strings $\astringofliterals_1, \dots, \astringofliterals_i$ such that each $\astringofliterals_k$, $1\le k\le i$, contains the literals assigned at level $i$.
	We define the order $<$ on string of integers as the lexicographic order on
	strings on integers.
	For any core state
	$\corestate{\astringofliterals}{\anoverapproximation}{\anunderapproximation}{\anaction}{}$
	we define
	$\depth{\corestate{\astringofliterals}{\anoverapproximation}{\anunderapproximation}{\anaction}{}}$
	as the string $2,\depth{\astringofliterals}$ if $\anaction\neq B$, and
	$0,\depth{\astringofliterals}$ if $\anaction= B$.
	We consider that any control
	state
	$\controlstate{\continueinstruction}{\astringofliterals}{\anoverapproximation}{\anunderapproximation}{\anaction}{}$
	is such that
	$\depth{\controlstate{\continueinstruction}{\astringofliterals}{\anoverapproximation}{\anunderapproximation}{\anaction}{}}=1$,
	and any state $s$ that is a terminal state is such that
	$\depth{s}=3$.
	
	We then define an order on the gap between over-approximation and
	under-approximation, which in general is
	$\anoverapproximation\setminus\anunderapproximation$.
	We define the functions $ove$ and $und$.
	For any state $s$, if $s$ is
	$\corestate{\astringofliterals}{\anoverapproximation}{\anunderapproximation}{\anaction}{}$ or
	$\controlstate{\continueinstruction}{\astringofliterals}{\anoverapproximation}{\anunderapproximation}{\anaction}{}$
	then $ove(s)=\anoverapproximation$ and $und(s)=\anunderapproximation$,
	otherwise $ove(s)=\emptyset$ and $und(s)=\literalsof{\atoms{\aprogram}}$.
	For two sets of literals $\asetofliterals$ and $\asetofliterals'$, we say that
	$\asetofliterals<\asetofliterals'$ if
	$\asetofliterals'\subseteq\asetofliterals$.
	
	We write $\lexlower$ to denote the lexicographic composition of orders.
	Then we define our order on states as follows.
	For any two states, $s<s'$ iff
	$(ove(s)\setminus und(s),\depth{s})\lexlower(ove(s')\setminus und(s'),\depth{s'})$.
	The relations on $\depth{s}$ and $ove(s) \setminus und(s)$ are
	clearly partial orders. Hence the obtained lexicographic order is also a
	partial order.
	We are now going to show that any edge $(s,s')$ in $\{\mathit{Oracle}\}
	\setunion\bigcup_{x\in S}x $ such that $s$ is reachable from the
	initial state is such that $s<s'$ and $s\neq s'$.
	Assume that a state $s$ is
	reachable from the initial state and the rule $\succeedrule$,
	$\failrule_{\underapproxaction{}}$ or $\failrule_{\chunkapproxaction{}}$
	applies to $s$ so as to create the edge $(s,s')$. Then by Lemma
	\ref{lemma:struct3}, $s<s'$ and $s\neq s'$.
	So, indeed, any edge $(s,s')$ in $\{\mathit{Oracle}\}
	\setunion\bigcup_{x\in S}x $ such that $s$ is reachable from the
	initial state is also such that $s<s'$ and $s\neq s'$.
	As a consequence, since the relation $<$ on states is a partial order and there
	is only a finite amount of ordered elements, there is no infinite path, and
	hence no cycle among the reachable elements of
	$(\statesof{\atoms{\aprogram}},
	\{\mathit{Oracle}\}
	\setunion\bigcup_{x\in S}x
	)$.
\end{proof}

\subsection{Proof of Theorem \ref{prop:general}}
By Lemmas \ref{lemma:finite} and \ref{lemma:acyclic}, the graph
$\agraph=(\statesof{\atoms{\aprogram}},
\{\mathit{Oracle}\}
\setunion\bigcup_{x\in S}x
)$ is finite and no cycle is reachable from the initial state.
Assume a state $\corestate{\astringofliterals}{\anoverapproximation}{\anunderapproximation}{\anaction}{}$
is terminal in $\agraph$; this is impossible since if no other rule applies then
$\succeedrule$ applies. Similarly, assume a state
$\controlstate{\continueinstruction}{\astringofliterals}{\anoverapproximation}{\anunderapproximation}{\anaction}{}$
is reachable and terminal in $\agraph$. Either
$\anoverapproximation=\anunderapproximation$ and $\mathit{Terminal}$ applies,
or $\anoverapproximation\neq\anunderapproximation$ and, by Lemma
\ref{lemma:struct1},
$\anunderapproximation\subset\anoverapproximation$ so one of the rules of the
nonempty set $\{\overapproxrule,\underapproxrule,\chunkapproxrule\}
\setintersection \bigcup_{x\in S}x$ applies. In both cases a rule applies, which
is a contradiction.

Therefore, the terminal state is $\terminalstate{\astringofliterals}$ for some
$\astringofliterals$. 
Hence, as to end the proof of the theorem we now study the
type of state that can actually be terminal.
Assume that $\terminalstate{\asetofliterals}$ is the terminal state reachable from the initial state. 
Either it was reached by a
transition justified by $\failrule_{\overapproxaction}$ and, by Lemma 
\ref{lemma:struct2}, in any state
$\corestate{\astringofliterals}{\asetofliterals}{\anunderapproximation}{\overapproxaction}{}$
from which this transition may have originated
holds $\anaim_\atypeofmodel=\asetofliterals$, or it was reached by a
transition justified by $\mathit{Terminal}$ and, by Lemma \ref{lemma:struct1},
in any state
$\controlstate{\continueinstruction}{\astringofliterals}{\asetofliterals}{\asetofliterals}{\anaction}{}$
from which this transition may have originated holds
$\asetofliterals\subseteq\anaim_\atypeofmodel\subseteq\asetofliterals$,
hence $\anaim_\atypeofmodel=\asetofliterals$.

\end{document}